\documentclass[11pt,leqno]{article}
\usepackage{amssymb,amsfonts}
\usepackage{amsmath,amsthm,amsxtra}
\usepackage[all]{xy}
\usepackage{color}
\usepackage{verbatim}

\newcommand\bigcheck[1]{#1 \raise1ex\hbox{$\hspace{-1ex}{}^\vee$}}
\newcommand\sucheck[1]{#1 \raise0.5ex\hbox{$\hspace{-1ex}{}^\vee$}}

\newcommand{\alphaparenlist}{
  \renewcommand{\theenumi}{\alph{enumi}}%
  \renewcommand{\labelenumi}{(\theenumi)}%
}

\newcommand{\romanparenlist}{
  \renewcommand{\theenumi}{\roman{enumi}}%
  \renewcommand{\labelenumi}{(\theenumi)}%
}

\newcommand{\cycl}{{\rm cycl}}

\newcommand{\even}{\mathop{\rm even \, }}

\newcommand{\odd}{{\rm odd}}
\newcommand{\ord}{{\rm ord}}

\renewcommand{\sl}{s\ell}

\newcommand{\C}{\mathcal{C}}

\newcommand{\R}{\mathcal{R}}
\newcommand{\V}{\mathcal{V}}
\newcommand{\W}{\mathcal{W}}

\newcommand{\CC}{\mathbb{C}}
\newcommand{\NN}{\mathbb{N}}

\newcommand{\ZZ}{\mathbb{Z}}

\newcommand{\fg}{\mathfrak{g}}

\renewcommand{\tilde}{\widetilde}

\makeatletter
\renewcommand\section{\@startsection {section}{1}{\z@}%
                                   {-3.5ex \@plus -1ex \@minus -.2ex}%
                                   {2.3ex \@plus.2ex}%
                                   {\normalfont\large\bfseries}}
\renewcommand\subsection{\@startsection{subsection}{2}{\z@}%
                                     {-3.25ex\@plus -1ex \@minus -.2ex}%
                                     {0ex \@plus .0ex}%
                                     {\normalfont\normalsize\bfseries}}

\@addtoreset{equation}{section}
\makeatother

\newtheorem{theorem}{Theorem}[section]

\newtheorem{lemma}[theorem]{Lemma}
\newtheorem{corollary}[theorem]{Corollary}
\newtheorem{proposition}[theorem]{Proposition}
\newtheorem{conjecture}[theorem]{Conjecture}
\newtheorem*{lemma*}{Lemma}

\theoremstyle{remark}
\newtheorem{remark}[theorem]{Remark}
\newtheorem{example}[theorem]{Example}

\makeatletter
\def\@maketitle{\newpage
 \null
 \vskip 2em
 \begin{center}%
  \vskip 3em
  {\Large\bf \@title \par}%
  \vskip 1.5em
  {\normalsize
   \lineskip .5em
   \begin{tabular}[t]{c}\@author
   \end{tabular}\par}%
  \vskip 2em

 \end{center}%
 \par
 \vskip 2.5em}
\makeatother

\newcommand{\mb}[1]{{\mathbb #1}}

\renewcommand{\epsilon}{\varepsilon}

\definecolor{light}{gray}{.9}

\begin{document}

\title{On classification of Poisson vertex algebras}

\author{Alberto De Sole\thanks{Dip. di Matematica, 
Universit\`a di Roma
``La Sapienza'', 00185 Roma, Italy.\,\,\,~~desole@mat.uniroma1.it~~~~\,\,
Supported
in part by Department of Mathematics, M.I.T. and by PRIN and AST grants},\,\,\,
~~ Victor G. Kac
\thanks{Department of Mathematics, M.I.T.,
Cambridge, MA 02139, USA.\,~~kac@math.mit.edu\,\,~~~~Supported in part by NSF 
grants~~
}~~\,\,
and Minoru Wakimoto\thanks{~~wakimoto@r6.dion.ne.jp\,\,~~~~Supported in part 
by Department of Mathematics, M.I.T.}}

\maketitle

\begin{center}
\emph{Dedicated to Vladimir Vladimirovich Morozov on his centennial}
\end{center}

\vspace{2pt}

\begin{abstract}
\noindent 
We describe a conjectural classification of Poisson vertex algebras of CFT type
and of Poisson vertex algebras in one differential variable
(= scalar Hamiltonian operators).\\

\noindent{Transformation Groups 15 (2010), no. 4, 883-907}
\end{abstract}

\section{Introduction}

\label{sec:intro}

Recall that a \emph{Poisson vertex algebra} (PVA) is a unital
commutative associative algebra $\V$ with a derivation~$D$,
endowed with a $\lambda$-bracket $\{ \,{} . {}_\lambda . {} \,\} : \V
\otimes \V \to \CC [\lambda] \otimes \V $, which satisfies the
axioms of a Lie conformal algebra, and the $\lambda$-bracket is
related to the product by the left Leibniz rule:
\begin{equation}
  \label{eq:0.1}
  \{ a_\lambda bc \} = \{ a_\lambda b \} c + b \{ a_\lambda c\}\,.
\end{equation}
In this case one says that the differential algebra $\V$ is endowed
with a \emph{Poisson $\lambda$-bracket}.

Recall, for completeness, that a \emph{Lie conformal algebra} is a
$\CC [D]$-module, endowed with a $\lambda$-bracket, which
satisfies the following three axioms \cite{K}:

\begin{eqnarray*}
  \begin{array}{ll}
\mbox{(sesquilinearity)}& \{ D a_\lambda b \}
    = -\lambda \{ a_\lambda b \}\, , \, \{ a_\lambda D b \}
    = (D + \lambda) \{ a_\lambda b\}\\[1ex]
\mbox{(skewcommutativity)} & \{ b_\lambda a \}
    = -{}_{\leftarrow}\!\{ a_{-\lambda -D}b \}\\[1ex]
\mbox{(Jacobi identity)} & \{ a_\lambda \{ b_\mu c \}\} 
     - \{ b_\mu \{ a_\lambda c \}\} 
     =  \{\{ a_\lambda b \}_{\lambda +\mu} c \} \, .
  \end{array}
\end{eqnarray*}
The left arrow in the second axiom means that $D$ is moved to
the left. Extension of these definitions to the super case is straightforward,
using the usual sign rule.

Recall that the left Leibniz rule and skewcommutativity imply the
right Leibniz rule \cite{DK}:
\begin{equation}
  \label{eq:0.2}
  \{ bc_\lambda a \} = \{ b_{\lambda +D} a \}_\rightarrow
    c + \{ c_{\lambda +D}a\}_\rightarrow b \, ,
\end{equation}
where the right arrow means that $D$ is moved to the right.

The awkward name ``Poisson vertex algebra'' comes from the fact
that it arises as a quasiclassical limit of a family of vertex
algebras \cite{DK} in the same way as a Poisson algebra arises as
a quasiclassical limit of a family of associative algebras.

Note that PVA is a local counterpart of a Coisson (=chiral
Poisson) algebra, defined in \cite{BD}.  Also PVA can be obtained
as a formal Fourier transform of a local Poisson bracket \cite{BDK}, which
plays an important role in the theory of infinite-dimensional
integrable Hamiltonian systems.  In fact, as
demonstrated in \cite{BDK}, the language of Poisson vertex
algebras is often more convenient and transparent than the
equivalent languages of local Poisson brackets, used in the book \cite{FT}, 
or of Hamiltonian operators, used in the book \cite{D}.

In the present paper we shall discuss the problem of
classification of Poisson $\lambda$-brackets on the
algebra of differential polynomials
\begin{displaymath}
  \R_\ell = \CC [u^{(n)}_i | i = 1,\ldots ,\ell \, ; \, 
  n \in \ZZ_+]
\end{displaymath}
in $\ell$ differential variables $u_i$, where the derivation
$D$ is defined in the usual way:  
$D u^{(n)}_i =
u^{(n+1)}_i$, $ n \in \ZZ_+$. As usual, we shall write $f'$ in place of
$D f$, in particular,
$u_i$, $u_i'$, $u_i''$,... shall often replace 
$u^{(0)}_i$, $u^{(1)}_i$, $u^{(2)}_i$,.... In the super case one considers the
algebra $\R_{\ell , m}$ of differential polynomials in $\ell$ even
differential variables and $m$ odd ones.

It is clear that, like in the Poisson algebra
case, a $\lambda$-bracket on $\R_\ell$ is uniquely determined by the
$\lambda$-brackets $\{ u_{i \lambda} u_j \}$, $ i,j =1,\ldots ,\ell$,
due to sesquilinearity and the left and right Leibniz rules.
It is explained in \cite{BDK} that, like in the Poisson algebra case,
the necessary and sufficient conditions for validity of PVA
axioms is skewcommutativity for each pair $u_i,u_j$ and Jacobi
identity for each triple $u_i,u_j,u_k$.  Like in the Poisson
algebra case, there is an explicit formula for the
$\lambda$-bracket of any $f$, $g \in \R_\ell$ in terms of $\lambda$-brackets 
of differential variables \cite{DK}:
\begin{equation}
  \label{eq:0.3}
  \{f_\lambda g \} 
  = \sum_{\substack{1 \leq i,j \leq \ell\\ m,n \in \ZZ_+}}
    \frac{\partial g}{\partial u^{(n)}_j} (D + \lambda)^n
    \{ u_{i_{D + \lambda}} u_j \}_\rightarrow
    (-D -\lambda)^m \frac{\partial f}
{\partial  u^{(m)}_i}\, .
\end{equation}

It turns out to be  more natural to consider an 
\emph{algebra of differential functions} extension 
$\tilde{\R}_\ell$ of $\R_\ell[x]$, that is a domain
$\tilde{\R}_\ell$, containing $\R_\ell[x]$, such that all partial derivatives
$\frac{\partial}{\partial u^{(n)}_i}$ extend to commuting
derivations of $\tilde{\R}_\ell$ and only finitely many functions
$\frac{\partial f}{\partial u^{(n)}_i}$ are non-zero for each $f
\in \tilde{\R}_\ell$.  Then 
$D$ extends to $\tilde{\R}_\ell$ by the formula
$D = \frac{\partial}{\partial x}+
\sum_{\substack{1 \leq i \leq \ell\\ n \in \ZZ_+}}u^{(n+1)}_i
\frac{\partial}{\partial u^{(n)}_i}$ and
formula~(\ref{eq:0.3}) extends the
$\lambda$-bracket from $\R_\ell$ to $\tilde{\R}_\ell$, making the
latter a PVA as well \cite{BDK}.  An element $f$ of $\tilde{\R}_\ell$ is
called a \emph{quasiconstant} (resp. a \emph{constant}) 
if $\frac{\partial f}{\partial u^{(n)}_i}=0$ for all $i$ and $n$ (resp.
if, in addition, $\frac{\partial f}{\partial x}=0$). We denote by $\C$
the subalgebra of all constants. 

Recall that, given a Poisson $\lambda$-bracket on 
$ \tilde{\R}_\ell$,
the associated Hamiltonian operator is the matrix 
$H=(H_{ij}(D))$, where $H_{ij}(D)= 
 \{ u_{j_{\,D}} u_i \}_\rightarrow$.
Conversely, the $\lambda$-bracket can be reconstructed from $H$ via 
\begin{equation}
\label{eq:0.4}
\{ u_{i_{\lambda}} u_j \} = H_{ji}(D+\lambda)(1).
\end{equation}
Also, given $\int h dx \in\tilde{\R}_\ell\ /D\tilde{\R}_\ell$, the 
corresponding system of Hamiltonian equations is $\frac{du}{dt}=
\{h_{\lambda}u\}|_{\lambda=0}
(=H \frac{\delta\int h dx}{\delta u})$ \cite{BDK}. 

Recall that the skewcommutativity of the 
$\lambda$-bracket is equivalent to the skewadjointness of $H$\cite{BDK}.
Hence, unlike in the Poisson algebra situation, the $\lambda$-bracket of
a function $f \in \tilde{\R}_\ell$ with itself can be non-zero.
In fact, the skewcommutativity axiom is equivalent to the
relation:
$\{ f_\lambda f \} = \sum_{j\, odd}
    (D +2 \lambda)^j F_j\, , \quad F_j \in \tilde{\R}_\ell$. 
Thus, even the cases of one or two differential variables
are already highly non-trivial.  Apart from a conjecture, stated
at the end of the introduction, we shall be concerned only with these two
cases.

In the case $\ell =1$ we have $  \R_1 = \CC [u,u',u'',\ldots ]$, and, 
according to the above remarks, 
a skewcommutative $\lambda$-bracket on $\tilde{\R}_1$ is determined by
\begin{equation}
  \label{eq:0.5}
  \{ u_\lambda u \} = \sum^N_{\substack{j=1\\j \mbox{\, odd}}}
    (D + 2 \lambda)^j f_j \, , \quad f_j \in \tilde{\R}_1\,.
\end{equation}
Here $N$ is a positive odd integer , called the \emph{order} of
the $\lambda$-bracket, provided that $f_N \neq 0$. Note that the Jacobi 
identity for the triple $u,u,u$ holds in the case when all the $f_j$ in 
(\ref{eq:0.5})
are quasiconstants. Such a $\lambda$-bracket is called a
\emph{quasiconstant coefficient} Poisson $\lambda$-bracket.

For an arbitrary Poisson $\lambda$-bracket on $\tilde{\R}_1$
the Jacobi identity for the triple $u,u,u$ gives a very
complicated system of PDE on the functions $f_j$.
In order to state our first result on the structure of these
functions, define the \emph{differential order} of $f \in
\tilde{\R}_1$, denoted by $\ord (f)$, as the maximal $m \in
\ZZ_+$, such that $\frac{\partial f}{\partial u^{(m)}} \neq 0$,
if $f$ is not a quasiconstant, and as $-\infty$ if $f$ is a quasiconstant.  
Define the \emph{level} $m$ of
the $\lambda$-bracket (\ref{eq:0.5}) of order~$N$ by
~$m=\max_j \{ j+ \ord (f_j) \}$. Note that $m$ is a positive integer 
if the $\lambda$-bracket is not a quasiconstant coefficient one.

\begin{theorem}
  \label{th:0.1}
The possible values~$m$ of the level of a non-quasiconstant coefficient 
Poisson 
$\lambda$-bracket (\ref{eq:0.5}) of order ~$N$ are $\frac{1}{2}(N-1) \leq m 
\leq 2N+1$, $m \neq 2N$, $m \neq \frac{1}{2}(N+1)$ if $N\equiv -1\mod 4$, and
 $m \neq \frac{1}{2}(N-1),\, \frac{1}{2}(N+3)$ if $N\equiv 1\mod 4$.
\end{theorem}

The notion of a level in the equivalent language of Hamiltonian operators
was considered by I.~Dorfman \cite{D}, who obtained our upper estimate
of the level by a different method. 
The classification of Hamiltonian operators of order  $N=1, 3$ and $5$,
obtained in 
\cite{V}, \cite{GD}, \cite{A}, \cite{AV}, \cite{O}, \cite{D}, \cite{M},\cite{C}, 
and some further calculations
lead to the following conjecture.  
\begin{conjecture}
\label{con:0.2}
With the exception of level $m=1$ in the case $N=3$,
the only possible values of the level of a non-quasiconstant coefficient 
Poisson $\lambda$-bracket of order $N$ are $m=N, N+1$, or $N+2$. 
(Examples below show that all these values of $m$
do occur, with the exception of $m=2$ in the case $N=1$.)
\end{conjecture}
 
This conjecture holds for $N \leq 11$, but for $N>11$ we can prove only that
$m < 2N-6$ by a more detailed analysis of 
the Jacobi identity (\ref{eq:1.1} ) for the triple $u, u, u$. 
(The proof of Theorem
\ref{th:0.1} uses only the highest total degree in $\lambda$ and $\mu$ 
term in (\ref{eq:1.1}).)\\   

From the conformal field theory (CFT) point of view, the most
interesting PVA are those which are obtained as a quasiclassical
limit from a family of vertex algebras of CFT type, which we shall
call the \emph{PVA of CFT type}.  By definition, this is an
algebra of differential polynomials $\R_{\ell }$ with $\ell$
differential variables $L, W_1, \ldots ,W_{\ell -1}$, endowed with a
of $\lambda$-bracket, satisfying the axioms of PVA, and such that

\romanparenlist
\begin{enumerate}
\item
$\{ L_\lambda L \} = (D +2\lambda) L +
  c\lambda^3$, where $c$ is a constant:

\item 
$\{ L_\lambda W_j \} = (D +\Delta_j \lambda) W_j$, \, $j=1,...,\ell -1$.
\end{enumerate}
Property~(i) says that the differential variable~$L$ generates the
Virasoro PVA (with central charge $12c$), while property~(ii)
says that $W_j$ is a \emph{primary element of conformal weight}
$\Delta_j$.  
Note that the augmentation ideal of $\R_{\ell}$ is
a Poisson ideal iff $c=0$.  Hence the simplicity of the PVA
$\R_{\ell }$ implies that $c \neq 0$.  
Important examples of PVA of CFT type are provided by classical W-algebras
$\W^k(\fg,f)$, associated to a simple Lie algebra $\fg$ and its nilpotent 
element $f$ (see e.g. \cite{DK}).
We prove the following theorem.

\begin{theorem}
  \label{th:0.3}
\alphaparenlist 
\begin{enumerate}
\item 
Let $\R_2$ be endowed with a PVA structure, generated
by a Virasoro differential variable $L$ 
with $c \neq 0$ 
and a primary differential variable ~$W$ of integer conformal weight~$\Delta>2$,
such that $\{W_\lambda W \} \neq 0$.
Then this PVA is isomorphic to one of the classical $W$-algebras 
$\W^k (\fg ,f)$, where $
 \fg$ is a simple Lie algebra of rank~2 and $f$ is a principal
 nilpotent element of $\fg$.

\item 
Let $\R_{1,1}$ be endowed with a super PVA structure, generated by an even
Virasoro differential variable L with $c \neq 0$ and an odd primary
differential variable $W$ of conformal weight $\Delta \in \frac12 + \NN$,
such that $\{W_\lambda W \} \neq 0$.
Then this PVA is isomorphic to the Neveu--Schwarz
  super PVA, namely, $\Delta =3/2$ and
  \begin{displaymath}
    \{ W_\lambda W \} =L +2 \lambda^2 c \, .
  \end{displaymath}
\end{enumerate}

\end{theorem}

This theorem supports the following conjecture.

\begin{conjecture} 
 \label{con:0.4}

Let $A=\CC[D]L \oplus \CC[D]W_1 \oplus ... \oplus \CC[D]W_{\ell -1}
\oplus \CC C$
be a $\CC[D]$-module with $D C=0$. Endow $S(A)$ with a Poisson
$\lambda$-bracket, for which 
$\{L_\lambda L\}=(D +2\lambda)L +\lambda^3C$
and the $W_j$ are primary of conformal weight $\Delta_j \in \NN$ (so that
$S(A)/(C-c)$ is a PVA of CFT type). Assume that $A$ does not contain proper 
non-zero $\CC[D]$-submodules $I$, such that $IS(A)$ is a PVA ideal 
of $S(A)$.
Then $S(A)/(C-c)$ is isomorphic to a 
classical $W$-algebra $\W^k (\fg,f)$, where $\fg$ is a simple Lie algebra
(including the 1-dimensional one), $f$  is its nilpotent element
and $c=-k(x|x)$ (here $f,x$ are elements of an $sl_2$ triple, such that 
$[x,f]=-f$).

\end{conjecture}

One can state a similar conjecture in the super case.\\

We prove Theorem \ref{th:0.1} in Section 1 and Theorem \ref{th:0.3} 
in Section 2. 

In Section 3 we discuss the problem of classification of scalar
Hamiltonian operators 
of arbitrary (odd) order $N$ (i.e. the case $\ell =1$), 
up to contact transformations. 
Recall that such a classification for 
$N\leq 5$ was obtained in a series of papers by Vinogradov, Gelfand-Dorfman, 
Astashov, Mokhov, Olver, and Cooke \cite{V}, \cite{GD}, \cite{A}, \cite{AV}, 
\cite{M},\cite{O}, \cite{C}.
We introduce the following new family of compatible
Hamiltonian operators of order $N=2n+3\geq 3$:
$H^{(N,0)}= D^2 \circ (\frac{1}{u}\circ D)^{2n} \circ D$.
We prove in Section 4 that these operators are Hamiltonian and compatible
(i.e. any their linear combination with constant coefficients is Hamiltonian).
Furthermore, in Section 3 we introduce a sequence of Hamiltonian operators
$H_{[N,c(x)]}$ 
of order $N\geq 7$, depending on a linear quasiconstant $c(x)$.

Our main observation is that any Hamiltonian operator of order $N\geq 7$ can 
be taken by a contact transformation to one of the following three types:

(1) a skew-adjoint differential operator with quasiconstant coefficients,

(2) a linear combination with constant coefficients of the operators 
$H^{(n,0)}, \,\,3\leq n \leq N$,

(3) the operators $H_{[N,c(x)]}$, where $N\geq 9$ and $c''(x)=0$,
 
(4) a ``small'' family of exceptional Hamiltonian operators.

We checked that this is indeed true for $7\leq N\leq 13$, and in Section 3 we
exibit in each of these cases the operators of type (4). 
The strategy of the proof is the same as in the above mentioned papers,
but the use of the machinery of Poisson vertex algebras considerably 
simplifies calculations. First, using Conjectures \ref{con:0.2} and
\ref{con:3.4} on the level and the leading coefficient, one shows that
by a contact transformation the leading coefficient can be made equal 1.
After that, using contact transformations that keep the leading coefficient
being 1, one reduces the Hamiltonian operator to a canonical form.
  
Remarkably, it turns out that for $N=13$ the set of operators of 
``exceptional'' type (4) is empty, i.e. 
any Hamiltonian operator of order $N=13$ can be taken by a contact 
transformation to an operator of type 
(1), (2) or (3)! We conjecture that the same holds for all  $N>13$.

In Section 3 we also analyse the hierarchies of integrable Hamiltonian
equations, obtained by the Lenard-Magri scheme \cite{Ma},\cite{BDK}
from a compatible pair
of Hamiltonian operators, which we call the bi-Hamiltonian integrable
equations, for one of our compatible pairs. On the basis of this analysis we 
state Conjecture \ref{con:3.15}
on classification of all scalar bi-Hamiltonian integrable equations.

We would like to thank A. Mikhailov, O. Mokhov and V. Sokolov
for enlightening discussions and correspondence. 


\section{Proof of Theorem~\ref{th:0.1}}
\label{sec:1}

The Jacobi identity for the $\lambda$-bracket (\ref{eq:0.5})
reads:
\begin{equation}
\label{eq:1.1}
  \{ u_\lambda \{u_\mu u\}\} - \{ u_\mu \{u_\lambda u \}\}
     = \{\{ u_\lambda u \}_{\lambda +\mu}u\}\, .
\end{equation}
Substituting (\ref{eq:0.5}) in (\ref{eq:1.1}) and using
(\ref{eq:0.3}), we obtain a polynomial equation in $\lambda
,\mu$ and the $\frac{\partial f_j}{\partial u^{(i)}}$.  The
highest total degree in $\lambda $ and $\mu$ in this equation is
$2N +m$.  Equating to $0$ this term, we obtain, after dividing
by $2^N f_N$:
\begin{displaymath}
  \sum^{\min \{ N,m \}}_{\substack{j=1\\j\mbox{\, odd}}} F_{N+m,j} 
(\lambda ,\mu)
    \frac{\partial f_j}{\partial u^{ (m-j)}}=0\, .
\end{displaymath}
Here the polynomials $F_{n,j} (u,v)$ for $1 \leq j \leq n$,
$j$~odd, are as follows:
\begin{displaymath}
  F_{n,j}(u,v)=u^{n-j}(v-w)^j + v^{n-j}(w-u)^j + w^{n-j}(u-v)^j\,,
\end{displaymath}
where we let $w=-(u+v)$.  Hence Theorem~\ref{th:0.1} follows
immediately from the following proposition.

\begin{proposition}
  \label{prop:1.1}

Let $N$ be a positive odd integer and $m=2N$ or $m \geq 2N+2$.
Then the collection  of polynomials $S_{N,m}: =
\{ F_{N+m,j} (u,v)\}_{1 \leq j \leq N,\, j\, \odd}$ is linearly independent.

\end{proposition}

The proof of the proposition is based on the following lemma.

\begin{lemma}
  \label{lem:1.2}
\alphaparenlist
\begin{enumerate}
\item 
$\left( \frac{\partial^2}{\partial u^2} 
     + \frac{\partial^2}{\partial v^2} 
         -\frac{\partial^2}{\partial  u\partial v} \right)
          F_{n,j}  =(n-j)(n-j-1) F_{n-2,j} + 3j (j-1)
          F_{n-2,j-2}$ (the RHS is zero if $n \leq 2$).

\item
$4 (u^2 + uv +v^2) F_{n,j} = 3F_{n+2,j} + F_{n+2,j+2}$.

\item 
If $n$ is divisible by $3$, then the polynomial $u^2+uv+v^2$
does not divide~$F_{n,1}$.

\end{enumerate}
\end{lemma}

\begin{proof}
(a) is straightforward.  (b) is obtained, using that $4 (u^2 + uv
+ v^2)$ can be written in the following three forms:   
\begin{displaymath}
  3u^2 + (u+2v)^2 = 3v^2 + (2u+v)^2 = 3 (u+v)^2 + (u-v)^2\, .
\end{displaymath}
Then we rewrite  the LHS of~(b), using consecutively for each of
the three summands these three forms.

(c) is proved by noting that $u^2+uv+v^2$ vanishes at $u=1$,
$v=\omega:=e^{2\pi i/3}$, while $F_{n,1} (1,\omega)=\omega-\omega^2 \neq 0$.

\end{proof}

\begin{corollary}
  \label{cor:1.3}

\alphaparenlist
\begin{enumerate}
\item 
If the collection of polynomials $S_{N,m}$ is linearly
independent, then the collection of polynomials $S_{N,m+2}$ is
linearly independent.

\item 
If $ N + m +2$ is divisible by~$3$ and the  collection
$S_{N,m}$ is linearly independent, then the  collection
$S_{N+2,m}$ is linearly independent.

\end{enumerate}

\end{corollary}

\begin{proof}
Applying to a linear dependence of elements from $S_{N,m+2}$, the
operator $\frac{\partial^2}{\partial u^2} +
\frac{\partial^2}{\partial v^2} -\frac{\partial^2}{\partial u
  \partial v}$ and using
Lemma~\ref{lem:1.2}(a), we obtain a linear dependence between the
polynomials $\{ (n-j) (n-j-1) F_{N +m,j} + 3j (j-1)
F_{ N+m,j-2}$, where $n = N+m+2$ and $1 \leq j \leq
N$, $j$~odd\}.  But this set is linearly independent if the set
$S_{N,m}$ is, proving~(a).

In order to prove (b), multiplying all elements from $S_{N,m}$ by
$u^2+uv+v^2$, we obtain, due to Lemma~\ref{lem:1.2}(b), a linearly
independent set of polynomials $S= \{ 3F_{n+2,j} + F_{n+2,j+2}$,
where $n=N +m$, $1 \leq j \leq N$, $j \, \odd \}$.  Since 
by Lemma~\ref{lem:1.2}(c), the
polynomial $F_{n+2,1}$ is not divisible by $u^2 + uv+v^2$,
we conclude that the set $S \cup \{
F_{n+2,1}\}$ is linearly independent, which implies that the set
$S_{N+2,m}$ is linearly independent.

Now we can complete the proof of Proposition~\ref{prop:1.1} by
induction on~$N$.  If $N=1$, then $S_{N,m}$ consists of one
non-zero polynomial.  Due to Corollary~\ref{cor:1.3}(a) it
suffices to prove the proposition when $m=2N$ and $m=2N+3$.
Note that, by the inductive assumption, $S_{N-2,2N}$ is linearly
independent, hence by Corollary~\ref{cor:1.3}(b), $S_{N,2N}$ is
linearly independent.  Similarly, by the inductive assumption,
$S_{N-2,2N+3}$ is linearly independent, hence by
Corollary~\ref{cor:1.3}(b), $S_{N,2N+3}$ is linearly independent.

\end{proof}

\begin{remark}
  \label{rem:1.4}

If $m <2N$  or $m=2N+1$, then the set of polynomials $S_{N,m}$ is
linearly dependent.  Indeed, we can view any of the polynomials
$F_{N+m,j}$ as an element of the set $S_{N+m}$ of polynomials of
the form $(u-v)(v-w)(w-u) f(u,v,w)$
where~$f$ is a symmetric polynomial in $u,v,w$, considered $\mod
(u+v+w)$, of degree~$N+m-3$.  Hence $\dim S_{N+m}$ equals the
number of partitions of $N+m-3$ in a sum of $2$'s and $3$'s.
But the latter number is smaller than $\dim S_{N,m} = (N+1)/2$ for 
the considered values of~$m$.

\end{remark}

\section{Proof of Theorem~\ref{th:0.3}}
\label{sec:2}

\begin{lemma}
  \label{lem:2.1}

Let $\V$ be a PVA of CFT type with all conformal weights positive.
Then an element~$P$ of~$\V$ is primary iff it is a polynomial in
$W_1, \ldots ,W_\ell$.

\end{lemma}

\begin{proof}

The ``if'' part is clear by the Leibniz rule~(\ref{eq:0.1}).  In
order to prove the ``only if'' part, note that we have by
induction on~$m$:
\begin{equation}
  \label{eq:2.1}
  \{ L_\lambda L^{(m)} \} = \lambda^{m+3} c + 
    (\mbox{lower powers of\,\,} \lambda)\, .
\end{equation}
Writing, as usual, $L_\lambda = \sum_{n \in
  \ZZ_+}\frac{\lambda^n}{n!} L_{(n)}$, we obtain a sequence of
derivations $L_{(m)}$ of $\V$ (due to (\ref{eq:0.1})).  Due to
(\ref{eq:2.1}), we have, in particular:
\begin{equation}
  \label{eq:2.2}
  L_{(n)} L^{(m)} = \delta_{n,m+3} \alpha c \mbox{\,\, for \,\,}
     n \geq m+3\, , 
\end{equation}
where $\alpha$ is a non-zero constant.

Write the element~$P$ as a polynomial in $W^{(n)}_j$, $j=1,
\ldots, \ell$, $n \in \ZZ_+$, with coefficients in $\CC [L,L',
L'',\cdots]$.  If one of these coefficients, say~$f$, is not
constant, let~$m$ be the maximal integer for which $L^{(m)}$
occurs in~$f$.  Then, due to (\ref{eq:2.2}), $L_{( m+3)}f =
\alpha c \frac{\partial f}{\partial L^{(m)}} \neq 0$.  Hence~$f$
is not primary.

Likewise, we have
\begin{displaymath}
  L_{(n)} W^{(m)}_j = \delta_{n,m+1} \Delta_j \mbox{\,\, if \,\,}
     n \geq m+1\, .
\end{displaymath}
Hence, if $m$ is maximal such that $W^{(m)}_j$ occurs for
some~$j$ in a polynomial~$P \in \CC [W^{(n)}_j | j=1,\ldots
,\ell\, \, n \in \ZZ_+]$, we have:
$L_{(m+1)} P= \sum_j \Delta_j \frac{\partial P}{\partial W^{(m)}_j}$.
Hence $P$ can be primary only if $P \in \CC [W_1,\ldots ,W_\ell]$.
\end{proof}

Recall that $L_{(0)}=D$ and $L_{(1)}$ is a diagonalizable
operator on a PVA~$\V$ of CFT type.  If $L_{(1)}P = \Delta P$,
$\Delta \in \CC$, one says that~$P$ has \emph{conformal
  weight}~$\Delta$ and lets $\Delta = \Delta_P$.  One has:
\begin{displaymath}
  \Delta_{PQ}=\Delta_P +\Delta_Q \, , \quad \Delta_{P'}
  =\Delta_P+1\, ,
\end{displaymath}
if $P$ and $Q$ are eigenvectors of $L_{(1)}$.  Moreover, the
coefficient of $\lambda^j$ in $\{ P_\lambda Q \}$ has conformal
weight $\Delta_P +\Delta_Q -j-1$.  In other words, all the
summands in $\{ P_\lambda Q\}$ have conformal weight $\Delta_P +
\Delta_Q -1$ if we put $\Delta_\lambda =1$.

Now let $\V$ be a PVA as in Theorem~\ref{th:0.3} and assume that
$\Delta \in \frac12 \ZZ_+$, $\Delta >1$.  It follows from the
skewcommutativity of the $\lambda$-bracket and the properties of
the conformal weight that
\begin{equation}
  \label{eq:2.3}
  \{ W_\lambda W \} = \sum^N_{\substack{j=1\\j \mbox{\,\, odd}}}
      (D + 2 \lambda)^jP_j \, , \quad P_N \neq 0\, ,
\end{equation}
where $N$ is a positive odd integer, $1 \leq N \leq 2 \Delta -1$
and $P_j$ has conformal weight $2\Delta -j-1$.  Hence we may
write:
\begin{equation}
  \label{eq:2.4}
  \{ L_\lambda P_j \}= ( D + (2\Delta -j-1)\lambda )P_j  + 
\sum^{2\Delta -j}_{k=2} \lambda^k Q_{j,k}\, ,
\end{equation}
where $\Delta_{Q_{j,k}} = 2\Delta -j-k$.

The following lemma is straightforward.

\begin{lemma}
  \label{lem:2.2}

The Jacobi identity for the triple $L,W,W$ is equivalent to the
following equations:
\begin{equation}
\label{eq:2.5}
  Q_{j,2k}  = \binom{j+2k}{j} P'_{j+2k},\,\,\, \, Q_{j,2k+1}= 
(2\Delta \binom{j+2k}{j}- \binom{j+2k+1}{j}) P_{j+2k}\, . 
\end{equation}

\end{lemma}

First, we assume that $\Delta$ is an integer $>2$, and we shall prove the 
claim (a) of the Theorem.
Denote by $a_j$ the coefficient of $L^{(2\Delta -3-j)}$ in $P_j$ and
by $b_j$ the coefficient of $W^{(\Delta-1-j)}$ in $P_j$, and let
$g=P_{2\Delta-1}$.  From the conformal weight considerations it
follows that $a_j$, $b_j$ and $g$ are constants (depending on $c$).

\begin{lemma}  
\label{lem:2.3}

If $g=0$ and $b_{\Delta -1}=0$, then $\{ W_\lambda W \}=0$.

\end{lemma}

\begin{proof}

By Lemma~\ref{lem:2.1}, the only primary elements in $\V$ are
  polynomials in~$W$. Since the conformal weight of $\{ W_\lambda
  W \}$ is $2\Delta -1$, we conclude that it may contain only
  $W^\epsilon$, where $\epsilon =0$ or~$1$.  Thus, by the
  conditions of the lemma, all~$P_j$ are not primary elements,
  unless they are~$0$.  Now by downward induction, beginning with
  $g=P_{2\Delta -1} =0$, we show, using (\ref{eq:2.5}) that all $
 P_j$ are primary, hence zero, and that all~$Q_{j,k}=0$.

\end{proof}

Introduce the following two polynomials in $x$ and $y$:
\begin{displaymath}
  G_1 (x,y)=\sum^{2\Delta -3}_{\substack{j=1\\j \mbox{\,\,odd}}}
    a_j x^{j}y^{2\Delta -3-j}\, , \quad 
  F_1 (x,y)=\sum^{\Delta -1}_{\substack{j=1\\j \mbox{\,\,odd}}}
    b_j x^j y^{\Delta -1-j} \, , 
\end{displaymath}
and put
\begin{displaymath}
  G (x,y) = G_1 \left( \frac{x-y}{2} \, , \, \frac{x+y}{2}\right)
  \, , \quad 
  F (x,y) = F_1  \left( \frac{x+y}{2} \, , \,
    \frac{x-y}{2}\right)\, .
\end{displaymath}
We obviously have:
\begin{equation}
  \label{eq:2.6}
  G (y,x)=-G(x,y)\, , \quad F(y,x)=(-1)^\Delta F(x,y)=-F(-x,-y)\, .
\end{equation}
Then the $W^{(r)}$ terms for $r \in \ZZ_+$ of the $WWW $ Jacobi
identity $\{ W_\lambda \{ W_\mu W \}\} - \{ W_\mu \{ W_\lambda W
\}\} = \{\{ W_\lambda W \}_{\lambda +\mu}W \}$ give the following
identity if we let $x=\lambda \, , \, y=\mu \, , \, z=D+\lambda +\mu$:
\begin{equation}
\label{eq:2.7}
  (x-(\Delta -1)(y+z)) G(y,z) + \cycl
   = 2F (x,-y) F (x+y , -z)+\cycl\, , 
\end{equation}
where ``$\cycl$'' means that we add two terms obtained from the
first one by cyclically permuting $x$, $y$, $z$.

The constant term of the $WWW$ Jacobi identity gives the
following identity, multiplied by the constant~$g$:
\begin{equation}
  \label{eq:2.8}
  x^{2\Delta -1}F (x+y,y)-y^{2\Delta -1}F (x+y,x)=(-1)^\Delta
     (x+y)^{2\Delta -1} F (x,-y)\, .
\end{equation}
Finally, the constant term of the $LWW$ identity gives
\begin{equation}
  \label{eq:2.9}
 c G (x,y)=48g \left( \Delta 
    \frac{x^{2\Delta -1}-y^{2\Delta-1}}{(x+y)^2} \, - \,  
    \frac{x^{2\Delta} -y^{2\Delta}}{(x+y)^3} \right) \,  .
\end{equation}
In particular,
\begin{equation}  
\label{eq:2.10}  
cG (x,0)=48g (\Delta -1)x^{2\Delta -3}\, .
\end{equation}
We also have
\begin{equation}
  \label{eq:2.11}
  F (0,x)=Ax^{\Delta -1} \mbox{\,\, for some constant\,\,}A \, .
\end{equation}
Letting $z=0$ in (\ref{eq:2.7}) and plugging (\ref{eq:2.9}),
(\ref{eq:2.10}) and (\ref{eq:2.11}) in it, we obtain:
\begin{align}
\label{eq:2.12}
   \frac{24gc (\Delta -1)}{(x+y)^2}&((\Delta -1)(x^{2\Delta
    -1}- y^{2\Delta -2} ) (x+y)^2 \tag{2.12}\\
  & + (xy^{2\Delta -3} -x^{2\Delta -3}y)(x+y)^2 -
    \Delta (x+y)(x^{2\Delta -1}-y^{2\Delta -1})+ x^{2\Delta} 
        - y^{2\Delta})\notag\\
   &  = A (x^{\Delta -1}+y^{\Delta -1} -(x+y)^{\Delta -1})
        F (-x,y)\, . \notag
\end{align}

First, we prove that $g=0$ if $\Delta$ is an odd integer $>3$.
In the contrary case, letting $y=0$ in (\ref{eq:2.8}), we get:
$F (x,0) = F (0,x)=0$.  Hence, letting $z=0$ in (\ref{eq:2.7}),
we get:
\begin{displaymath}
  (x-(\Delta -1)y) G (y,0) - (y-(\Delta -1)x) G (x,0) =
      (\Delta -1) (x+y) G (x+y)\, .
\end{displaymath}
Substituting (\ref{eq:2.10}) in this formula, we obtain (after
canceling $48g c (\Delta -1)$):
\begin{displaymath}
  (\Delta -1)(x^{2\Delta-2}-y^{2\Delta -2}) - x^{2\Delta-3} y
     + xy^{2\Delta -3} = (\Delta -1)x^{2\Delta -2}-(\Delta -2)
       x^{2\Delta -3}y + (\Delta -3)x^{2\Delta -4} y^2 +\cdots\, ,
\end{displaymath}
which is impossible if $\Delta >3$.

Next, we consider the case $g=0$. Since $\Delta \neq 2$, it follows from
 (\ref{eq:2.12}) that $A=0$, hence, by (\ref{eq:2.11}),
\addtocounter{equation}{1}
\begin{equation}  
\label{eq:2.13}  
F (0,x)=0 .
\end{equation}  
Also, it follows from (\ref{eq:2.6}) that
\begin{equation}  
\label{eq:2.14}  
F (-x,x)=0.
\end{equation}
Since $g=0$, by (\ref{eq:2.9}), 
$G(x,y)=0$, hence (\ref{eq:2.7}) becomes:
\begin{equation}
\label{eq:2.15}
  F (x,-y)F (x+y, -z) + F (y,-z) F(y+z,-x) + F(z,-x)F(x+z,-y) =0 \, .
\end{equation}
We will show, using the above three equations, that $F(x,y)=0$, which 
implies, in particular, that
$b_{\Delta -1}=0$, and we can apply Lemma \ref{lem:2.3}.

For this, consider the function $\varphi(t)=F(x,tx)/x^{\Delta -1}$.
This is a polynomial in t.

\begin{lemma}  
\label{lem:2.4}
The polynomial $\varphi (t)$ has the following properties:

\begin{list}{}{}
\item (i)~~$\varphi (a) =0$, $a \neq 0 \Rightarrow \varphi (a^{-1})=0$,

\item (ii)~~$\varphi (-2) \varphi (3) =0$,

\item (iii)~~$\varphi (a) =0$, $a \neq 0 \Rightarrow \varphi
  (a^{-2}) \varphi (a+a^{-1})=0$.
\end{list}

Properties (ii) and (iii) imply that $\varphi (t) =0$.

\end{lemma}

\begin{proof}

(i) follows from (\ref{eq:2.6}) since $F (x,y)$ is a homogeneous polynomial.  
In order to prove~(ii), let $y=-2x$, $z=x$ in (\ref{eq:2.14}) and
use (\ref{eq:2.12}) and (\ref{eq:2.13}).  Similarly, (iii)~is
proved by letting $x=az$, $y=a^{-1}z$ in (\ref{eq:2.14}) and
using~(i).

Next, by (ii), $\varphi (a_1) =0$ for some $a_1$, such that
$|a_1| >1$.  Since $|-a^2_1|>|a_1|$ and $|a_1 + a^{-1}_1 | >
|a_1|$, it follows from~(iii) that $\varphi (a_2)=0$ for some
$a_2$, such that $|a_2|>|a_1|$, etc.  Thus the polynomial
$\varphi (t)$ has infinitely many zeroes, hence equals zero.

\end{proof}

Lemma \ref{lem:2.4} completes the proof of (a) in the case $g=0$. 
Now consider the case $g \neq 0$.  Then identity~(\ref{eq:2.8})
holds and also we obtain from (\ref{eq:2.12}), letting there $y=1$:
\begin{align}
\label{eq:2.16}
  \mbox{the polynomial\,\,} & H(x)=(\Delta -3)x^{2\Delta -2}(x+1)-
x^{2\Delta -3}+x^3-(\Delta -3)x(x+1)
\tag{2.16}\\
      & 
\mbox{\,\, is divisible by the polynomial\,\,}
        f(x)= (x+1)^{\Delta -1}- x^{\Delta -1}-1\, .\notag
\end{align}
\addtocounter{equation}{1}
By the above discussion, we may assume that $\Delta$ is an even integer.
Now we will show that (\ref{eq:2.16}) is impossible for all positive even 
integers $\Delta > 6$.


In order to show that (\ref{eq:2.16}) does not hold for these
$\Delta$, note that $f(x)=f(-x-1)$, hence the divisibility of
$H(x)$ by $f(x)$ implies the divisibility of $H_1 (x) := -(x+1) H
(-x-1)$ by $f(x)$.  Next, we have:  $H_1 (x) \equiv H_0(x) \mod
(f)$, where $H_0 (x) = (\Delta -3)x (x+1) x^{2\Delta -2}-
x^{2\Delta -2} +2 (\Delta -3) (x^{\Delta +1} +x^\Delta)
-2x^{\Delta -1} + (x+1)^4 + (\Delta -3) x (x+1)^2 + (\Delta -3) x
(x+1)-1$.  Then $P(x) := H_0(x) - xH (x) = 2 (\Delta -3)
x^{\Delta +1} + 2 (\Delta -3)x^\Delta - 2x^{\Delta -1} + 2
(\Delta -1)x^3 + 2 (2\Delta -3)x^2+ 2(\Delta -1)x$ is divisible
by $f(x)$ iff $H_0 (x)$ is.  Now assume that $\Delta >6$.  Then,
dividing $P(x)$ by $f(x)$, we obtain the remainder $Ax^{\Delta
  -3}+$~lower degree terms, where $A \neq 0$.  Thus, $H(x)$ is
not divisible by $f(x)$ for $\Delta $~even~$>6$.

We conclude the proof of (a) by a direct computation in cases
$\Delta =3$, $4$ and~$6$.  For example, in the case of $\Delta
=3$, from the conformal weight considerations, we have: $\{ W_\lambda W \}
  = (D+ 2\lambda) P_1 + (D+ 2\lambda)^3 P_3 +
  (D + 2\lambda)^5 P_5$, where $P_1 = \alpha L^2 + \beta
  L'' +\gamma W'$, $P_3 = \delta L$, $P_5=\epsilon$, and
  $\alpha$, $\beta$, $\gamma$, $\delta$, $\epsilon$ are constants,
  not all equal to zero.  Then the $LWW$ Jacobi identity determines
  these five constants up to a non-zero common factor, and after
  rescaling, we get:
  \begin{displaymath}
    P_1 = 2^8L^2 +2^3 \cdot 3 cL''\, , \, 
    P_3 = 2^3 \cdot 5cL \, , \, 
    P_5 =  c^2\, .
  \end{displaymath}
The $WWW$ Jacobi identity then automatically holds, and we obtain
the classical $W$-algebra $\W^k (\sl_3,f)$.

Similar, but more complicated computations give for $\Delta =4$:
$ \{ W_\lambda W \} = \sum^7_{\substack{j=1\\j\,\odd}} 
    (D +2\lambda)^j P_j$, where
\begin{align*}
  P_1 &= 2^{11} \cdot 3^{2} L^3 +  2^{6} c L^{'2}
    + 2^{7} \cdot 29 c LL'' + 2^4 \cdot 3 c^2 L^{(4)}
    +2^4 \cdot 7\sqrt{2} LW+ 2 \sqrt{2} cW''\, , \\
  P_3 &= 2^{6} \cdot 7^2 cL^2 +2^5 \cdot 7c^2 L'' + 
2 \cdot 3 \cdot \sqrt{2} cW \, , \, 
  P_5 = 2^4 \cdot 7 c^2 L \, , \, P_7 = c^3\, .
\end{align*}
This is the classical $W$-algebra $\W^k (sp_4,f)$.

Finally, for $\Delta =6$ we get:
$\{ W_\lambda W \} = \sum^{11}_{\substack{j=1\\j\,\odd}} 
      (D + 2\lambda)^j P_j$, where 
$\deg_1 P_j = 11-j$ with $\deg_1L=2$, $\deg_1 W=6$,
$\deg_1 D =1$, $\deg_1 c =0$ and $\deg_2 P_j =5$ with
$\deg_2 L = \deg_2 c =1$, $\deg_2 W=3$, $\deg_2 D =0$.
The explicit formulas for $P_j$ are too long to be reproduced
here.  This is the classical $W$-algebra $\W^k (G_2,f)$.

The proof of (b) is similar, but simpler.  We use again
Lemmas~3.1 and~3.2.    In this case the polynomials $P_j$ do not
contain the terms, linear in $W^{(m)}$, $m \in \ZZ_+$, since
$\Delta \in \frac12 + \ZZ$, hence the linear terms of $P_j$ are
linear combinations of $L^{(m)}$, $m \in \ZZ_+$ and~$1$.  Denote
the coefficient of $L^{(2\Delta -3-j)}$ in $P_j$ by $a_j$, and
let $g=P_{2\Delta -1}$.  Then $a_j =0$ if $j$ is odd.  As before,
let
\begin{displaymath}
  G_1 (x,y) = \sum^{2\Delta -3}_{\substack{j=0\\j \even}}
    a_jx^jy^{2\Delta -3-j}\, , \qquad
  G (x,y) = G_1 \left( \frac{x-y}{2} \, , \, \frac{x+y}{2} \right)\,.
\end{displaymath}
As before, computing the $W^{(r)}$-terms of the WWW Jacobi
identity, we get:
\begin{equation}
  \label{eq:2.17}
  ((\Delta -1)(z-y)+x) \,\, G (z,-y)+  \cycl =0 \, .
\end{equation}
The constant term of the LWW identity gives:
\begin{equation}
  \label{eq:2.18}
  G (x,y) = 48g \sum^{2\Delta -3}_{\substack{j=0\\j\odd}}
    (-1)^j (j+1) \left(\Delta - \frac{j+2}{2}\right)
      x^{2\Delta -3-j}y \, .
\end{equation}
Plugging (2.18) in (2.17) and computing the coefficient of $z^2$,
we get:
\begin{displaymath}
  48g ((3\Delta -6)(xy^{2\Delta -5} + x^{2\Delta -5} y)) -
     (\Delta -1) (\Delta -3) (x^{2\Delta -4}-y^{2\Delta -4})=0\, .
\end{displaymath}
But this is impossible if $\Delta \geq 5/2$ and $g \neq 0$.
Hence $\{ W_\lambda W \}=0$ if $\Delta \geq 5/2$, proving~(b).

\begin{example}
  \label{ex:2.5}

The differential algebra $\R_2$ with $W$ of conformal weight~$4$
and the $\lambda$-bracket
\begin{displaymath}
  \{ W_\lambda W \} = (D +2 \lambda) (\alpha L^3 +\beta LW)
\end{displaymath}
is a PVA with central charge $c=0$ for any values of $\alpha,
\beta  \in \CC$, however only for the value $\alpha / \beta = 2^{6} \cdot 3^2
\cdot \sqrt{2}/7$  this PVA is a member of a family of
PVA, depending on arbitrary central charge~$c$.
\end{example}

\section{On classification of Poisson $\lambda$-brackets of 
arbitrary order~$N$ in one differential variable}
\label{sec:3}

In this section we study Poisson $\lambda$-brackets on the algebra of
differential functions 
$\V =\tilde{\R}_1$.  The inclusion of $x$ allows us to consider contact
transformations, see \cite{A}, \cite{AV}, \cite{M}, which
preserve the order of the $\lambda$-bracket (but do not preserve
translation invariance, i.e. independence of the coefficients of $x$).  

A {\em contact transformation} of the differential algebra~$\V$
is a transformation of the form:
\begin{equation}  
\label{eq:3.1}
  x=\varphi (y,v,v_y)\, , \, u = \psi (y,v,v_y)\, ,
\, D_x =\frac{1}{\varphi'}D_y,\tag{3.1}
\end{equation}
such that the following conditions hold:
\begin{displaymath}
  \frac{\partial \varphi}{\partial v_y} \psi' =
     \frac{\partial \psi}{\partial v_y} \varphi',\,\,\,\, \varphi'
\hbox{\, and\,\,}
      \rho \varphi' := \frac{\partial \psi}{\partial v}
        \varphi' - \frac{\partial \varphi}{\partial v}\psi'
\hbox{ \,are invertible elements of} \,\V .
\end{displaymath}
Note that the Jacobian of the transformation 
$(y,v,v_y)\mapsto (x,u,u_x)$
equals $\rho^2$.
An example of a contact transformation is the Legendre transformation:
$\varphi =v',\,\psi=yv'-v$ (for which $\rho=-1$).
The contact transformations are precisely all automorphisms of the algebra
$\V$, which leave invariant the contact form 
$\omega =du-u'dx$ up to multiplication by a function (the factor being
$\rho$), and also precisely those transformations which preserve the order 
of any Hamiltonian operator \cite{M}. 

\begin{example}
\label{ex:3.1}
The Legendre transformation takes the operator $D^N$ to the 
differential operator 
$T_N := \frac{1}{u''}(D \circ \frac{1}{u''})^N$ \cite{M}. 
It follows that
the operators $T_N$ for positive odd $N$ are Hamiltonian and compatible,
which answers the question, raised in \cite{C1}.
\end{example}

\begin{example}
\label{ex:3.2}
The contact transformation $\varphi =v,\,\,\psi=-y$ takes the translation
invariant Hamiltonian operator $D\circ (\frac{1}{u'}D)^{N-1}$ to
a quasiconstant coefficient (but not translation invariant) Hamiltonian 
operator $D^N +2xD +1$.
\end{example}
 
We will say that two $\lambda$-brackets on~$\V$ or two Hamiltonian operators
are {\it equivalent} if in an algebra of  
differential functions extension of $\V$ one of them can be
transformed to another by a contact transformation.  
The following result is well known. 
\begin{theorem}
  \label{th:3.3}
\cite{V}, \cite{GD}, \cite{A}, \cite{AV}, \cite{M}, \cite{O}
Any Hamiltonian operator in one function $u$
of order $N=1$ (resp. $N=3$) is equivalent to the following (unique) one:
\begin{displaymath}
   D  \quad (\hbox{resp.\,\,}     
D^3 + a(2uD+u') 
    \quad a \in \C)\, .
\end{displaymath}

\end{theorem}

Using Conjecture \ref{con:0.2} for order $N$ Poisson $\lambda$-brackets,
the first step in their classification is the following conjecture,
which we checked for $N\leq 11$.
 
\begin{conjecture}
\label{con:3.4}
Let $f_N$ be the leading coefficient of a Poisson $\lambda$-bracket on $\V$
of order $N$ (recall that $N$ is odd) and level $m=N+\epsilon$, where 
$\epsilon=1$ or $2$, and let $N_{\epsilon}=N+2\epsilon-3$.
Then
\begin{displaymath}
N_{\epsilon}f_N\frac{\partial ^2f_N}{\partial u^{(\epsilon) 2}}=
(N_{\epsilon} +1)(\frac{\partial f_N}{\partial u^{(\epsilon)}})^2.
\end{displaymath}
Equivalently: $f_N=\frac{a}{(u^{(\epsilon)}+b)^{N_{\epsilon}}}$,
where $a,b \in \V$ have differential order at most $\epsilon -1$.
\end{conjecture}

The following remark shows that Hamiltonian operators remain Hamiltonian under
contact transformations.
\begin{remark}
\label{rem:3.5}
Given an element $P=P(x,u,u',\ldots) \in \V$ (resp. a differential operator
$H$), denoted by $\tilde{P}$ (resp. $\tilde{H}$) the element 
(resp. differential
operator), obtained from $P$ (resp. $H$) by the substitution (\ref{eq:3.1}).
Then under the contact transformation (\ref{eq:3.1}), an evolution PDE
$\frac {du}{dt}=P$
gets transformed to $\frac {dv}{dt}=\frac{1}{\rho}\tilde{P}$, and, for
$\int hdx \in \V/D \V$, 
the variational derivative $\frac{\delta \int hdx}{\delta u}$ gets 
transformed to $\frac{1}{\rho\varphi'}\frac{\delta\int\tilde{h}dy}{\delta v}$.
A Hamiltonian (evolution) PDE
$\frac {du}{dt}=H\frac{\delta \int h dx}{\delta u}$, where $H$ is a 
Hamiltonian operator, gets transformed to
$\frac {dv}{dt}=H_{new}
\frac{\delta \int \tilde{h}dy}{\delta v}$, where
$H_{new}=\frac{1}{\rho}\tilde{H}\circ \frac{1}{\rho \varphi'}$ is again a
Hamiltonian operator \cite{M}.  
\end{remark}

The following remark shows how the first two coefficients of a
$\lambda$-bracket change under contact
transformations.
\begin{remark}
\label{rem:3.6}
A contact transformation takes the 
$\lambda$-bracket (\ref{eq:0.5})
to a $\lambda$-bracket of the form (\ref{eq:0.5}) with some coefficients $g_j
\in \V$, where $g_N=\tilde{f}_N/ \rho^2 \varphi'^{N+1}$. Furthermore,
$f_N=g_N=1$ if and only if the contact transformation has the form:
$\varphi=\varphi(y)$, $\psi =\varphi'(y)^{-(N+1)/2} v + f(y)$, and then
$g_{N-2}=\varphi'^2 \tilde{f}_{N-2} +\frac{1}{3}n(n^2-1) S(\varphi)$,
where $S(\varphi)=\varphi'''/\varphi' -3\varphi''^2/2\varphi'^2$ is the 
Schwarz derivative.
\end{remark}

A classification of Hamiltonian operators of order $N=5$ was 
obtained by Cooke \cite{C}, who showed that such an operator can be 
transformed either to a quasiconstant coefficient skewadjoint operator, 
or to a certain canonical form, depending on one parameter. This canonical 
form is not translation invariant, but it can be
slightly simplified, using the contact transformation with
$\varphi =e^{y}\, ,\, \psi =e^{-3y}v$, to make it translation invariant.
It turned out that, by the further contact transformation 
$\varphi =y, \psi = v^2/2$, the canonical form can be recast in a 
beautiful form, which can be easily generalized to arbitrary order $N$.
\begin{theorem}
    \label{th:3.7}
\alphaparenlist
\begin{enumerate}
\item 
Any Poisson Hamiltonian operator of order $N=5$ on~$\V$
is equivalent to either a quasiconstant coefficient skewadjoint 
operator or one of the following translation invariant
Hamiltonian operators ($b,c \in \C$):
\begin{equation}
\label{eq:can}
(D^2 -c^2)
\circ \frac{1}{u}\circ D \circ \frac{1}{u}(D^2 -c^2) \,+\,
bD (D^2 -c^2)\, .\tag{3.2}
\end{equation}
These two types of Hamiltonian operators are not equivalent.
The Hamiltonian operators (\ref{eq:can}), corresponding to parameters 
$(b,c)$ and $(b_1,c_1)$, 
are equivalent if and only if either $b\neq 0$ and $bc=\pm b_1c_1$, or 
$b=b_1=0$.
\item 
The (compatible for all $c \in \C$) Hamiltonian operators 
\begin{displaymath}
 H^{(5,c)}=(D^2 -c^2)\circ 
\frac{1}{u}\circ D\circ \frac{1}{u}(D^2 -c^2)
\hbox{\, and \,} K_c=D (D^2 -c^2)
\end{displaymath}
give rise to the Lenard-Magri scheme
\begin{equation}
\label{eq:3.2}
 K_c \xi_{j+1,c} = H^{(5,c)} \xi_{j,c} \, , \, j=0,1,2,\ldots ,\hbox{\, 
with \,}\xi_{0,c} = -\frac{2}{c^2}\,\,\hbox{if}\,\,c\neq 0,\xi_{0,0} = 
x^2\, ,\,\xi_{1,c} = \frac{1}{u^2}\, ,\tag{3.3}
\end{equation}
where, for $j\geq 1$, 
$\xi_{j,c}$ lie in $\V_0: =\C[u,u^{-1},u^{(n)};\,n\geq 1]$ and depend 
polynomially on c. This scheme produces an integrable hierarchy
of Hamiltonian evolutionary equations $\frac{du}{dt_j} =
K_c \xi_{j+1,c}$, $j=0,1,2,\ldots$, the first one being (after
rescaling):
\begin{equation}
\label{eq:3.3}
  \frac{du}{dt_0} = 
(\frac{u''}{u^3}-3 \frac{u'^2}{u^4}- \frac{c^2}{2}\, \frac{1}{u^2})'\,.\tag{3.4}
\end{equation}

\end{enumerate}

\end{theorem}

\begin{proof}
The proof of (a) consists of three steps (cf. 
\cite{AV}, \cite{M}, \cite{C}). First, one proves that the level
$m \leq 7$ (cf. Conjecture \ref{con:0.2}, which is proved for $N=5$).
Second, one shows, as in \cite{C}, that, by a contact 
transformation one can make the leading coefficient equal 1. 
(Here one uses Conjecture \ref{con:3.4}, which is proved for $N=5$.)
Third, one shows that any Poisson $\lambda$-bracket 
$\{u_{\lambda} u\}=\sum_{j=1,3,5}(D +2\lambda)^j f_j$ with $f_5=1$, 
either has $f_1$ and $f_3$
quasiconstant, or has the following $f_1$ and $f_3$,
where one can add to $u$ an arbitrary quasiconstant (this is 
Proposition 4.10 from \cite{C}, recast in terms of $\lambda$-brackets):

\begin{align}  
\label{eq:3.5}
f_1= & \frac{1}{4}c_1^2 +3c_1'' +2c_1\frac{u''}{u} -
c_1\frac{u'^2}{u^2}-4c_1'\frac{u'}{u}+\frac{1}{2}c_1c_2u+3c_2u''
-3c_2\frac{u'^2}{u}-3c_2''u +4\frac{u^{(4)}}{u}\tag{3.5}\\
       &- 12\frac{u'u'''}{u^2}+ 12\frac{u''^2}{u^2}-
24\frac{u'^{2}u''}{u^3}+ 21\frac{u'^{4}}{u^4},\,\,
f_3=  c_1+c_2u+ 12\frac{u''}{u}- 14\frac{u'^2}{u^2},  \notag
\end{align}  
where the $c_j=c_j(x)$ are quasiconstants, satisfying the following equation: 
\begin{equation}
\label{eq:c}
2c_1c_2'+c_2c_1'+4c_2'''=0.\tag{3.6}
\end{equation}
By Remark \ref{rem:3.5},
a contact transformation keeps the leading coefficient
to be 1 iff $\varphi=\varphi(y)$ and $\psi=\varphi'(y)^{-3}v+f(y)$, 
and this transformation takes the pair of quasiconstants 
$c_1,c_2$ to the pair of quasiconstants $\bar{c}_1,\bar{c}_2$, 
where  $\bar{c}_1=\varphi '^2 c_1(\varphi)+4S(\varphi)\,
,\bar{c}_2=c_2(\varphi)/\varphi '$. 
By such a transformation one can make $c_1=0$, so that $c_2(x)=Ax^2+Bx+C$,
where $A,B,C \in \CC$. We can make $C=0$, replacing $x$ by $x +const$.
Then, applying the transformation with $\varphi=y^{-1}$,
we make $A=0$ and keep $c_1=0$ since $S(\varphi)=0$, and if $B\neq 0$,
we can make $C=0$, replacing $x$  by $x+const$. Taking further $\varphi=e^y$,
we make both $c_1,c_2$ constant, and, finally, by the transformation
$\varphi =y, \psi = v^2/2$, the corresponding 
Hamiltonian operator is reduced to the 
form, described in (a), where $c=c_1,\,b=c_2$. The equivalence, stated in
(a) is immediate by the above remarks. All these computations
use Mokhov's transformation formula (see Remark \ref{rem:3.5}). For $N>5$
the strategy is the same, but the computational difficulties increase
exponentially.

In order to prove (b), let $\bar{H}_c=\frac{1}{u}\circ D
\circ \frac{1}{u}\circ (D^2 -c^2)$, $\bar{K}=D$,
so that $H^{(5,c)}=(D^2 -c^2)\bar{H}_c$, 
$K_c=(D^2-c^2)\bar{K}$.
Obviously, any solution to the Lenard-Magri scheme
\begin{equation} 
\label{eq:len}
D \xi_{j+1,c} = \bar{H}_c \xi_{j,c} \, , \, j=0,1,2,\ldots, \tag{3.7}
\end{equation} 
with $\xi_{0,c} = -\frac{2}{c^2}$ if $c\neq 0$, $\xi_{0,0} = 
x^2$, $\xi_{1,c} = \frac{1}{u^2}$, is a solutions to the Lenard-Magri scheme 
(\ref{eq:3.2}).
Note that $\bar{H}_c =D \circ H_c -
\frac{1}{2}K_c\xi_{1,c}$, where 
$H_c=\frac{1}{u^2} \circ (D^2 -c^2) +\frac{u'}{u^3}(D +c)
-(D +c)\circ \frac{u'}{u^3}$. Hence (\ref{eq:len}) becomes
\begin{equation}
\label{eq:part}
D \xi_{j+1,c}=D (H_c\xi_{j,c})-\frac{1}{2}\xi_{j,c}(K_c\xi_{1,c}).\tag{3.8}
\end{equation}
We construct a solution $\xi_{0,c},\ldots, \,\xi_{n,c}$ to (\ref{eq:len}),
such that $\xi_{j,c}\in \V_0$ for $j=1,\ldots,n$,
by induction on n. We already have it for $n=1$. If we have it for $n$,
then, it is also a solution to the Lenard-Magri scheme (\ref{eq:3.1})
with skew-adjoint operators, hence, by \cite{BDK}, Lemma 2.6, 
$\xi_{j,c}(K_c\xi_{i,c})\in D \V_0$
for all $i,j=1,\ldots,n$. In particular this holds for $j=n$ and $i=1$,
which implies that equation (\ref{eq:part}) has a solution 
$\xi_{j,c}\in \V_0$ for $j=1,\ldots,n+1$.
Using the recurrent formula (\ref{eq:part}), it is easy to prove by induction 
that the $\xi_{j,c}$ are linearly independent and, for $j\geq 1$, depend 
polynomially on $c$.

Letting $h_0=-\frac{2u}{c^2}$ if $c \neq 0$ and $=x^2u$ if $c=0$,
and $h_1 =-\frac{1}{u}$, so that 
$\xi_{j,c}=\frac{\delta \int h_j dx}{\delta u}$ for $j=0,1$,
by \cite{BDK}, Theorem 2.7 and Proposition 1.9, there exist
$\int h_2 dx,\int h_3 dx, \ldots \in \V_0/D\V_0$, such that 
$\xi_{j,c}=\frac{\delta\int h_j dx}{\delta u}$ for $j\geq 2$, 
hence the hierarchy, defined in (b), is integrable.
\end{proof}

Equation (\ref{eq:3.3}) appears in \cite{MSS} in a classification
of integrable evolution equations (see equation (4.1.26) there).

\begin{remark}
\label{rem:3.8}
The contact transformation $\varphi=y$, $\psi =v^2/2$ takes the Poisson
$\lambda$-brackets with $f_5=1$ and $f_1$, $f_3$ as in 
(\ref{eq:3.5}), to the Poisson $\lambda$-brackets 
$\{u_{\lambda} u\}_{(5,c_1,c_2)}=\sum_{j=1,3,5}(D +2\lambda)^j g_j$, where 
\begin{align*}  
g_1= & \frac{1}{u^6}(\frac{1}{4} c_1^2u^4 +c_1c_2u^6 -2c_1u^3u'' 
+6c_1u^2u'^2+3c_1''u^4-6c_2''u^6-8c_1'u^3u'-2u^3u^{(4)}+24u^2u'u'''\\
&+18u^2u''^2-144uu'^2u''+120u'^4),\,\,
g_3=\frac{1}{u^4}(c_1u^2+2c_2u^4+4uu''-12u'^2),\,\, 
g_5=\frac{1}{u^2}, 
\end{align*}  
and $c_i=c_i(x)$ satisfy (\ref{eq:c}). 
The corresponding Hamiltonian operator is, provided that $c_1(x)\neq 0$: 
\begin{displaymath}
H_{(5,c_1,c_2)}
=B_{(2,c_1,c_2)}^*\circ D \circ B_{(2,c_1,c_2)}\,,
\end{displaymath}
where $B_{(2,c_1,c_2)}=\frac{1}{u}D^2 +a(x) D -a'(x)-\frac{a''(x)}{a(x)u}$,
$c_1(x)=-8\frac{a''(x)}{a(x)}$, $c_2(x)=-2a(x)^2$, $a(x)$ is a non-zero
quasiconstant, and
\begin{displaymath}
H_{(5,0,c(x))}=H^{(5,0)}+\frac{1}{2}c(x)D^3+\frac{3}{4}c'(x)D^2,
\end{displaymath}
where $c'''(x)=0$.
If $c_2=0$, then $H_{(5,c_1,c_2)}$ is equivalent to $H^{(5,0)}$; 
otherwise it is equivalent 
to $H^{(5,c)}+\frac{1}{2}K_c$, where $8c^2=2c_2'^2-4c_2c_2''-c_1c_2^2$ 
(which is a constant due to (\ref{eq:c})).
\end{remark}

Motivated by Theorem \ref{th:3.7}, introduce the following skew-adjoint
differential operator of order $N=2n+3\geq 3$ and the leading coefficient
$u^{3-N}$:
\begin{equation}
\label{eq:3.6}
H^{(N,0)}= D^2 \circ (\frac{1}{u} D)^{2n} \circ D\,.
\tag{3.9}
\end{equation}

We show in the next Section that all the operators $H^{(N,0)}$ are 
Hamiltonian and compatible.
Note that two linear combinations of the form $H^{(N,0)}+ a_1 H^{(N-2,0)}+
a_2 H^{(N-4,0)}+...+a_n H^{(3,0)}$, where all $a_j \in \C$, 
are equivalent if and only if there exists 
a non-zero constant $s$, such that $a_j$ is replaced by $s^ja_j$ for all $j$.

It is straightforward to show that the contact transformation 
$\varphi =(e^{ay}-1)/a$, $\psi =e^{-by}v$, where $a=(n+1)c$, $b=(n+2)c$, 
$c\in \C$, takes the Hamiltonian operator $H^{(N,0)}$ to the Hamiltonian 
operator
\begin{displaymath}
H^{(N,c)}=(-1)^n(D -c)\circ (B^{(n,c)})^*\circ D \circ B^{(n,c)}\circ (D+c)\,,
\end{displaymath}
where 
\begin{displaymath}
B^{(n,c)}= \frac{1}{u}(D -c) \circ \frac{1}{u}\circ (D -2c)
\circ \frac{1}{u}\circ ... \circ \frac{1}{u}(D -nc)\,,
\end{displaymath}
$*$ stands for taking the adjoint differential operator, and $B^{(0,c)}=1$. 

The same arguments as in the proof of Theorem \ref{th:3.7}(a)
lead to a classification of Hamiltonian operators of order 7, 9 and 11.
We state here the results, omitting the detailed proofs.
\begin{theorem}
\label{th:3.9}
Any Hamiltonian operator of order 7 is equivalent either to a quasiconstant
coefficient skew-adjoint differential operator or to the operator 
$H_{(7,c(x))}+b^2 D^3$, where 
\begin{displaymath}
H_{(7,c(x))}=-B_{(3,c(x))}^*\circ D\circ B_{(3,c(x))},\,\, 
B_{(3,c(x))}=\frac{1}{u}D\circ \frac{1}{u}D^2 +
c(x)D-\frac{1}{2}c'(x), 
\end{displaymath}
and $c'''(x)=0$, $b \in \C$. 
%
%
%
%
%
%
These two types of Hamiltonian operators are not equivalent. The Hamiltonian
operators $H_{(7,c(x))}+b^2 D^3$ and $H_{(7,c_1(x))}+b_1^2 D^3$ are equivalent
if and only if $\alpha^2 c_1(x)= c(\alpha^3 x+\beta)$ and 
$\alpha^2 b_1 = \pm b$ for some
constants $\alpha \neq 0$ and $\beta$. Such a Hamiltonian operator is
equivalent to a linear combination of the operators $H^{(j,0)}$ if and only
if $c(x)=c\in \C$, 
and one has: $H_{(7,c)}= H^{(7,0)}+2c H^{(5,0)}+c^2 H^{(3,0)}$. 
\end{theorem}

\begin{remark}
\label{rem:3.10}
The compatible pair of Hamiltonian operators $H_{(7,c(x))}$, where 
$c'''(x)=0$, and
$D^3$ give rise to a Lenard-Magri scheme with 
$\xi_0 =1$ if $c''(x) \neq 0$ (resp. $\xi_0=c(x)$ if 
$c''(x)=0,\, c'(x) \neq 0$)
and $\bar{\xi}_0 =-\frac{1}{2} x^2$.  The
first Hamiltonian equations of the resulting hierarchy are (up to a constant
factor):

\begin{displaymath}
\frac{du}{dt_0} =H_{(7,c(x))} \xi_0= \left( 
\frac{u''}{u^3}- \frac{3u'^2}{u^4} -\frac{1}{4}\frac{(c(x)^2)''}{c''(x)}
\right)'\, \,\,(\hbox{resp.}\,\,=
\left(\frac{u''}{u^3}- \frac{3u'^2}{u^4} -\frac{3}{2}c(x)\right)'),
\end{displaymath}

$$
\begin{array}{rl}
\displaystyle{
\frac{du}{d\bar{t}_0} 
}&\displaystyle{
 =H_{(7,c(x))} \bar{\xi}_0
   = \left( \frac{u^{(4)}}{u^5}-15 \frac{u'u'''}{u^6} - 
      10 \frac{u''^2}{u^6} + 105 \frac{u''u'^2}{u^7}-
         105 \frac{u'^4}{u^8}\right.  
}\\[1ex]
& \displaystyle{
 + (2 c (x)-\frac{1}{4} c'' (x) x^2+\frac12 c'(x) x)
         \frac{u''}{u^3}
       - (\frac32 c'(x) x +6 c (x) - \frac{3}{4} c'' (x) x^2) \frac{u'^2}{u^4}  
}\\[1ex]
& \displaystyle{
       +5 c'(x) \frac{u'}{u^3} - \frac{5}{4}
           \frac{c''(x)}{u^2}
       -\frac{15}{16} c(x)^2  + \frac{9}{16}c (x) c'' (x) x^2 - \frac{9}{8} c (x) c'(x)x 
}\\[1ex]
& \displaystyle{
        \left.  + \frac{3}{64} c'' (x)^2 x^4- 
        \frac{3}{16} c' (x) c'' (x) x^3
          +\frac{3}{16} c' (x)^2 x^2 \right)'\,= D^3 \frac{\delta h_1}
        {\delta u} \, ,
}
\end{array}
$$
where
\begin{align*}
  h_1 &= a (x) u + \frac1u \left(c(x) -\frac18 c'' (x) x^2 
     + \frac14 c'(x)x\right) - \,\frac{u^{\prime 2}}{2u^5}\\
\noalign{\hbox{and}}\\
   a(x) &= \frac{x^2}{32} \left(-18 c(x)^2 - 6 c(x) c''(x) x^2
      + 20 c(x) c'(x)x -c'' (x)^2 x^4 + 5c' (x) c'' (x) x^3
         -7c'(x)^2 x^2\right)\, .
\end{align*}
We can show that this Lenard-Magri scheme produces an integrable
hierarchy of Hamiltonian equations. The first of these equations includes
equations (4.1.23) and (4.1.24) from \cite{MSS}, while the second seems to be 
new.
\end{remark}

For $N=2n+5 \geq 7$ and a quasiconstant $c(x)$, such that $c''(x)=0$,
introduce the following skew-adjoint differential operator of
order~$N$ with the leading coefficient $u^{3-N}$:
\begin{align*}
  H^{[N,c(x)]} &= (-1)^n (B^{[n+2,c(x)]})^*\circ D \circ
  B^{[n+2,c(x)]}\, ,\\
\noalign{\hbox{where}}\\
   B^{[n+2, c(x)]} &= \bigg(\frac1u D-c(x)\bigg)\ldots 
     \bigg( \frac1u D-nc(x)\bigg)
     \bigg(\frac1u D^2 + c(x) D-c'(x)\bigg)
\end{align*}
is a differential operator of order $n+2$.  By the same method,
as in Section~4, one shows that $H^{[N,c(x)]}$ is a Hamiltonian
operator.

For $N=7$ we have:  $H_{(7,-c(x)^2)}=  H^{[7,c(x)]}$ if $c''(x)=0$.

\begin{theorem}
\label{th:3.11}
Any Hamiltonian operator of order~9 is equivalent either to a
quasiconstant coefficient skew-adjoint differential operator,
or to the Hamiltonian operator $H^{[9,c(x)]} + a D^3$, where $c''(x)=0$,
$a \in \C$,  
or to the Hamiltonian operator  
$H_{(9,c(x))}
+a^2 H_{(5,0,c(x)/2)}+ b^3D^3$, where    
\begin{displaymath}
 H_{(9,c(x))}=
B^*\left(D\circ \frac{1}{u}\circ D \circ \frac{1}{u}D+c(x)D+\frac{1}{2}c'(x)
\right)B,
\end{displaymath}
\begin{displaymath}
B=\frac{1}{u}\circ D \circ \frac{1}{u}D^2+c(x)D-\frac{1}{2}c'(x), 
\end{displaymath}
$H_{(5,0,c(x))}$ is defined in Remark \ref{rem:3.8},
and $a,b \in \C$, $c'' (x) =0$.
These three types of Hamiltonian operators of order 9 are not equivalent.
The Hamiltonian operators, corresponding to the triples
$(c(x),a,b)$ and $(c_1(x),a_1,b_1)$ are equivalent if and only if
$\alpha c_1 (x) = c (\alpha^2x + \beta)$, $\alpha a_1 =\pm  a$, and
$\alpha b_1=\epsilon b$ for some constants $\alpha\neq 0$, $\beta$
and a cube root of unity $\epsilon$.
An operator $H_{(9, c(x))}$ is equivalent to a linear combination of the 
operators $H^{(j,0)}$ with constant coefficients
if and only if $c(x)=c \in \C$,  
and one has: $H_{(9,c)}= H^{(9,0)}+3c H^{(7,0)}+3c^2H^{(5,0)}+c^3H^{(3,0)}$. 
\end{theorem}

 \begin{remark}
\label{rem:3.11}
The compatible for $N \leq 9$ pair of Hamiltonian operators $H^{[N,c(x)]}$ 
and $D^3$ gives rise to a Lenard-Magri scheme with 
$\xi_0 =1$ and $\bar{\xi}_0 =x^2$. We can show 
that this Lenard-Magri scheme produces an integrable
hierarchy of Hamiltonian equations if $N=9$. 
The simplest equation of this hierarchy
is, up to a constant factor, the following equation of order 5, 
which seems to be new (here $c''(x)=0$, $c'(x)\neq 0$):
\begin{align*}
& \frac{du}{dt_0} = H^{[9,c(x)]} 1= \left( \frac{u^{(4)}}{u^5} -
    15 \frac{u'u'''}{u^6} - 10 \frac{u''^2}{u^6} + 105
      \frac{u''u'^{2}}{u^7}-105 \frac{u'^4}{u^8}\right.\\
    & + 20 c' (x) \frac{u'^2}{u^5}-5c'(x) \frac{u''}{u^4}
       +5c' (x)^2 \frac{1}{u^2} -20c (x) c' (x) \frac{u'}{u^3}\\
    &  + \left. 15 c (x)^2 \frac{u'^2}{u^4} - 5c (x)^2 \frac{u''}{u^3}
       -5c (x)^4 \right)'\, .
\end{align*}
(For $N=5$ we get in this way the 
integrable hierarchy in Theorem \ref{th:3.7} and for $N=7$ a special case of 
the integrable hierarchy of Remark \ref{rem:3.10}.) 
\end{remark}

\begin{theorem}
  \label{th:3.13}

Any Hamiltonian operator of order~$11$ is equivalent either to a
quasiconstant coefficient skew-adjoint differential operator, or
to a linear combination with constant coefficients of the
operators $H^{(j,0)}$ with $3 \leq j \leq 11$, $j$ odd, or to the
operator  

\begin{align*}
& H^{[11,c(x)]} 
   + a\left( \frac{1}{u^4} D^7 - 14 \frac{u'}{u^5} D^6
        + \frac{1}{3u^6}\left(-10c(x)^2u^4 - 8c' (x) u^3 -60uu'' 
    +285u'^2 \right)D^5 \right. \\
&    + \frac{5}{3u^7} \left( 10c (x)^{2}u^4 u'-
         10c (x) c' (x) u^5 + 12c' (x) u^3 u' 
         - 9u^2 u''' + 120uu'u''-225u'^3 \right) D^4\\
&   + \frac{1}{3u^8}\Big( 7c(x)^4 u^8 + 40c (x)^2 u^5 u''
        -120c (x)^2 u^4 u'^2+174c(x) c' (x) u^5  u' - 40c' (x)^2u^6 \\
&   + 48c' (x) u^4u''-192c'(x)u^3u'^2
          -18u^3u^{(4)}
    +  300 u^2u'u''' +210u^2 u''^2-2340uu'^2u''
    +2520 u'^4 \\
&   -\frac43 au^8 \Big) D^3
    + \frac{1}{3u^9}\Big( 42c(x)^3 c'(x)u^9 + 10c(x)^2u^6u'''
        -90c (x)^2u^5u'u'' + 120c(x)^2u^4u' \\ 
&   +81c(x) c' (x) u^6 u'' -243c (x) c'(x) u^5 u'^2 + 81c'(x)^2u^6u' + 12c'(x) u^5u''' 
    -144c' (x)u^4u'u'' \\
&   + 240c' (x)u^3u'^3-3u^4u^{(5)} +60u^3u'u^{(4)}
       +105u^3u''u'''
    - 585u^2u'^2u'''-810u^2u'u''^2 \\
&   + 3465uu'^3u''-2520u'^5\Big) D^2
    + \frac{7}{3u^5} c(x) c'(x) \left( 3c(x)c'(x)u^5 
        + u^2u''' -9uu'u''+12u'^3\right) D \\
&    \left. + \frac{7}{3u^5}c'(x)^2 \left( -3c(x)c'(x)u^5 
         -u^2u''' + 9uu'u''-12u'^3 \right)\right)\, ,
\end{align*}
where $ c'' (x)=0,\,  a\in \C$, or to the operator
$H_{(11, c(x))} +aH_{(5,0 ,2c(x) )}$, where
\begin{align*}
  H_{(11,c(x))} &= -B^*_{(11,c(x))} \circ D \circ
  B_{(11,c(x))}\,,\\
   B_{(11,c(x))} &= \left( \frac1u D \circ \frac1u D +
     \frac14 c(x) \right) \circ
    \left( \frac1u D \circ \frac1u D^2 + c(x) D -\frac12
     c'(x) \right)\, , 
\end{align*}
and $a\in \C,\,c'' (x)=0$, or to the operator 
$H_{[11, c(x)]}+aH_{(5,0 ,14c(x))}$,
$a\in \C,\,c''(x)=0$, where $H_{[11, c(x)]}$ is too long to be reproduced here 
(its computer printout takes 7 pages).

\end{theorem}

 \begin{remark}
\label{rem:3.14}
The compatible pairs of Hamiltonian operators 
($H_{(11,c(x))}$,  $H_{(5,0,2c(x))}$) and 
($H_{[11,c(x)]}$,  $H_{(5,0,14c(x))}$) 
give rise to Lenard-Magri schemes with 
$\xi_0=1$ and $\bar{\xi}_0 =c(x)$.
We conjecture that these two Lenard-Magri schemes produce integrable 
hierarches of Hamiltonian equations.
\end{remark}

Now we proceed to state our conjectures on classification of
Hamiltonian operators (in one differential variable~$u$) of
arbitrary (odd) order $N$.

\begin{conjecture}  
\label{con:3.12}
\alphaparenlist
\begin{enumerate}
\item 
Any Hamiltonian operator is equivalent to a Hamiltonian operator with the
leading coefficient 1. Note that for $N=2n+3\geq 3$ the latter is equivalent
to a Hamiltonian operator $H$ with the leading coefficient $u^{3-N}$ by the
contact transformation $x=y,\,u=v^{n+1}/(n+1)$. 
\item 
If $H$ is a translation invariant non-constant coefficient Hamiltonian 
operator of order $N=2n+3\geq 7$ with the leading coefficient 1, 
then after the contact transformation 
$x=y,\,u=v^{n+1}/(n+1)\,\,+const.$
it becomes either $H^{(N,c)}$ with $c \in \C$, 
or a linear combination with constant coefficients
of the operators $H^{(j,0)}$ with $3\leq j \leq N$, $j$ odd.
\item 
Any translation invariant Hamiltonian operator of order $N \geq 7$ is 
equivalent to either a quasiconstant coefficient skew-adjoint differential 
operator, or a linear combination of the operators $H^{(j,0)}$ with 
$3 \leq j \leq N$, $j$ odd.
\item 
For $N \geq 7$ the Hamiltonian operator $H^{(N,1)}$ is not compatible 
with any translation invariant Hamiltonian operator
other than a scalar multiple of itself.
\item 
Any Hamiltonian operator of order $N \geq 13$ is equivalent to either a 
quasiconstant coefficient skew-adjoint differential operator, or a 
linear combination with constant coefficients of the operators $H^{(j,0)}$ 
with $3 \leq j \leq N$, $j$ odd, or to the Hamiltonian operator 
$H^{[N,c(x)]}$, where $c''(x)=0$.
\item 
For $N \geq 11$ the Hamiltonian operator $H^{[N,c(x)]}$ with $c'(x)\neq 0$,
$c''(x)=0$ is compatible only with a constant multiple of itself.
\end{enumerate}
\end{conjecture}

We verified conjectures \ref{con:3.12} (a),(b),(d),(e) and (f) 
for $N\leq 13$, but were unable to prove (c) even for $N=7$.
Note that Conjecture \ref{con:3.12}
(a) follows from Conjectures \ref{con:0.2} and \ref{con:3.4},
and, conversely, it  implies these conjectures.  

\begin{remark}
\label{rem:3.15}
It follows from (\ref{eq:3.6}) that for $\xi_j: =\xi_{j,c=0}$, 
where $\xi_{j,c}$ are the same as in Theorem \ref{th:3.7}(b), we have:
\begin{displaymath}
D^3\xi_{j+n}=H^{(2n+3,0)}\xi_{j},\,\,\,j=0,1,2,\ldots,\,\,n=1,2,\ldots.
\end{displaymath}
In particular $\bar{\xi_j}:=\xi_{nj}$ give a solution to the Lenard-Magri
scheme $D^3 \bar{\xi}_{j+1} = H^{(2n+3,0)} \bar{\xi_{j}},\,\, 
j=0,1,\ldots$.
\end{remark}
\begin{remark}
\label{rem:3.16}
Let $H_{(5,ax)}$ denote the Hamiltonian operator, corresponding to
the $\lambda$-bracket $\{ u_\lambda u \}_{(5,c_1,c_2)}$ from
Remark 3.8 with $c_1=0$, $c_2=ax$, where $a$ is a non-zero
constant, and let $H_{(9,ax)}$ be the Hamiltonian operator
from Theorem \ref{th:3.11} with $c(x)=ax$.
Then the triple of Hamiltonian operators $H_{(9,ax)}$, $H_{(5,ax)}$,
$D^3$ is compatible.  Let $\xi_n$ be the sequence produced by the
Lenard-Magri scheme
\begin{displaymath}
  H_{(5,ax)}\xi_n = D^3 \xi_{n+1}\, , \quad \xi_0=x^2 \, .
\end{displaymath}
Then $H_{(9,ax)} \xi_0 = H_{(5,ax)}\xi_2$.

\end{remark}
In view of Conjecture \ref{con:3.12} and Remarks 
\ref{rem:3.15}, \ref{rem:3.16}, 
the following conjecture,
consistent with known classification results of general integrable equations
\cite{MSS}, seems natural. 
\begin{conjecture}  
\label{con:3.15}
Any integrable bi-Hamiltonian 
equation in $u$ is equivalent by a contact transformation to one, 
contained in either the linear hierarchy, or the KdV hierarchy, 
or the HD hierarchy, or the hierarchy 
defined in Theorem \ref{th:3.7}(b), or the hierarchies discussed in Remarks
\ref{rem:3.10}, \ref{rem:3.11} and \ref{rem:3.14}.
(For the definition of integrability of a Hamiltonian equation and 
the construction via the Lenard-Magri scheme of the KdV and HD 
hierarchies see e.g. \cite{BDK}.)  
\end{conjecture}  


\section{Compatible family of Hamiltonian operators $H^{(N,0)}(D)$}

Recall the definition \eqref{eq:3.6} of the operator $H^{(N,0)}(D)$, 
where $N\geq3$:
\begin{equation}\label{0327:eq1}
H^{(N,0)}(D)= D^2\circ B^{(N-3)}(D)\circ D
\,\,,\,\,\,\,\text{ where }
B^{(n)}(D):=\Big(\frac1u D\Big)^n\,.
\end{equation}
We denote by $\{\cdot\,_\lambda\,\cdot\}_N$ the corresponding $\lambda$-bracket,
which is given by
\begin{equation}\label{0327:eq2}
\{u_\lambda u\}_N
\,=\,
(\lambda+D)^2 B^{(N-3)}(\lambda+D)\lambda\,.
\end{equation}
\begin{theorem}\label{0327:th}
The operators $H^{(N,0)}(D),\, N\geq 3$ odd, are compatible Hamiltonian operators.
Namely, any linear combination
\begin{equation}\label{0327:eq9}
\{u_\lambda u\}
=
\sum_{\stackrel{N\geq 3,\text{ \emph{odd} }}{\text{(\emph{finite})}}}
\alpha_N\{u_\lambda u\}_N\,,
\end{equation}
with constant coefficients $\alpha_N$, is a Poisson $\lambda$-bracket.
\end{theorem}
\begin{lemma}\label{0327:lem1}
For every $m,n\in\mb Z_+$ we have
\begin{equation}\label{0327:eq3}
\begin{array}{l}
\big\{B^{(m)}(\lambda+D)1_{\lambda+\mu}u\big\}_{n+3} \\
=
-\lambda(\lambda+\mu+D)^2  B^{(n)}(\lambda+\mu+D) 
\Big(
B^{(m)}(\lambda+D)\frac1u
-(-1)^m B^{(m)}(\mu+D)\frac1u
\Big)\,.
\end{array}
\end{equation}
\end{lemma}
\begin{proof}
We prove equation \eqref{0327:eq3} by induction on $m\in\mb Z_+$.
For $m=0$ both sides of \eqref{0327:eq3} are zero.
For $m\geq0$
we have, by the definition \eqref{0327:eq1} of the operator $B^{(n)}(D)$,
\begin{equation}\label{0327:eq5}
\begin{array}{l}
\vphantom{\Big(}
\big\{B^{(m+1)}(\lambda+D)1_{\lambda+\mu}u\big\}_{n+3} 
=
\big\{\frac1u(\lambda+D)B^{(m)}(\lambda+D)1_{\lambda+\mu}u\big\}_{n+3} \\
\vphantom{\Big(}
=
- {\big\{u_{\lambda+\mu+D}u\big\}_{n+3}}_{\vphantom{\Big(}\!\!\!\!\!\!\to} 
\frac1{u^2}(\lambda+D)B^{(m)}(\lambda+D)1
- {\big\{B^{(m)}(\lambda+D)1_{\lambda+\mu+D}u\big\}_{n+3}}_{\vphantom{\Big(}\!\!\!\!\!\!\to} 
(\mu+D)\frac1u
\end{array}
\end{equation}
In the last identity we used sesquilinearity and the right Leibniz formula.
By the definition \eqref{0327:eq2} of the $\lambda$-bracket $\{\cdot\,_\lambda\,\cdot\}_{n+3}$,
the first term in the RHS of \eqref{0327:eq5} is equal to
\begin{equation}\label{0327:eq6}
\begin{array}{l}
- (\lambda+\mu+D)^2 B^{(n)}(\lambda+\mu+D)(\lambda+\mu+D)
\frac1{u}B^{(m+1)}(\lambda+D)1 \\
= -\lambda (\lambda\!+\!\mu\!+\!D)^2 B^{(n)}(\lambda+\mu+D)
\bigg(
B^{(m+1)}(\lambda+D)\frac1{u} 
+\Big(B^{m}(\lambda+D)\frac1u\Big)\Big((\mu+D)\frac1{u}\Big)
\bigg)\,.
\end{array}
\end{equation}
By inductive assumption, the second term in the RHS of \eqref{0327:eq5}
is equal to
\begin{equation}\label{0327:eq7}
-\lambda(\lambda+\mu+D)^2  B^{(n)}(\lambda+\mu+D) 
\bigg(
\Big(-B^{(m)}(\lambda+D)\frac1u\Big)\Big((\mu+D)\frac1u\Big)
+(-1)^m B^{(m+1)}(\mu+D)\frac1u
\bigg)\,.
\end{equation}
Combining \eqref{0327:eq6} and \eqref{0327:eq7}, we conclude that the RHS of \eqref{0327:eq5}
is equal to
$$
- \lambda(\lambda+\mu+D)^2 B^{(n)}(\lambda+\mu+D)
\bigg(
B^{(m+1)}(\lambda+D)\frac1{u}
+(-1)^m B^{(m+1)}(\mu+D)\frac1u
\bigg)\,,
$$
thus proving the claim.
\end{proof}
\begin{lemma}\label{0327:lem2}
For every $m,n\in\mb Z_+$ we have
\begin{equation}\label{0327:eq4}
\begin{array}{c}
-\frac1u
\Big((\lambda+D)  B^{(m)}(\lambda+D)\lambda\Big)
\Big((\mu+D)  B^{(n)}(\mu+D)\mu\Big)
+ \big\{u_\lambda B^{(n)}(\mu+D)\mu\big\}_{m+3} \\
= -\lambda^2\mu^2
B^{(n)}(\lambda+\mu+D)  B^{(m)}(\lambda+D)\frac1u \,.
\end{array}
\end{equation}
\end{lemma}
\begin{proof}
We prove equation \eqref{0327:eq4} by induction on $n\in\mb Z_+$.
For $n=0$ both sides of \eqref{0327:eq4} are equal to
$-\lambda^2\mu^2B^{(m)}(\lambda+D)\frac1u$.
For $n\geq0$, we have, by sesquilinearity and the left Leibniz formula,
\begin{equation}\label{0327:eq8}
\begin{array}{l}
\vphantom{\Big(}
\big\{u_\lambda B^{(n+1)}(\mu+D)\mu\big\}_{m+3}
= \big\{u_\lambda \frac1u(\mu+D)B^{(n)}(\mu+D)\mu\big\}_{m+3} \\
\vphantom{\Big(}
= - \frac1{u^2} \big\{u_\lambda u\big\}_{m+3} (\mu+D)B^{(n)}(\mu+D)\mu
+ \frac1u (\lambda+\mu+D) \big\{u_\lambda B^{(n)}(\mu+D)\mu\big\}_{m+3}\,.
\end{array}
\end{equation}
The first term in the RHS of \eqref{0327:eq8} is equal to
$$
- \frac1{u}
\Big((\lambda+D)^2B^{(m)}(\lambda+D)\lambda\Big)
\Big(B^{(n+1)}(\mu+D)\mu\Big)\,,
$$
and summing this to 
$$
-\frac1u
\Big((\lambda+D)  B^{(m)}(\lambda+D)\lambda\Big)
\Big((\mu+D)  B^{(n+1)}(\mu+D)\mu\Big)\,,
$$
we get
$$
- \frac1{u}
(\lambda+\mu+D)\frac1u
\Big((\lambda+D) B^{(m)}(\lambda+D)\lambda\Big)
\Big((\mu+D)B^{(n)}(\mu+D)\mu\Big)\,.
$$
Combining the above results we get, by the inductive assumption,
$$
\begin{array}{l}
-\frac1u
\Big((\lambda+D)  B^{(m)}(\lambda+D)\lambda\Big)
\Big((\mu+D)  B^{(n+1)}(\mu+D)\mu\Big)
+ \big\{u_\lambda B^{(n+1)}(\mu+D)\mu\big\}_{m+3} \\
=
\frac1{u}
(\lambda+\mu+D)
\bigg(
-\frac1u\Big((\lambda+D) B^{(m)}(\lambda+D)\lambda\Big)
\Big((\mu+D)B^{(n)}(\mu+D)\mu\Big) \\
\,\,\,\,\,\,\,\,\,\,\,\,\,\,\,\,\,\,\,\,\,\,\,\,\,\,\,\,\,\,\,\,\,\,\,\,\,\,\,\,\,\,\,\,\,\,\,\,\,\,\,\,\,\,\,\,\,\,\,\,\,\,\,\,\,\,\,\,\,\,\,\,
+ \big\{u_\lambda B^{(n)}(\mu+D)\mu\big\}_{m+3}
\bigg) \\
\vphantom{\Big(}
= -\lambda^2\mu^2
B^{(n+1)}(\lambda+\mu+D)  B^{(m)}(\lambda+D)\frac1u \,,
\end{array}
$$
as we wanted.
\end{proof}
\begin{proof}[Proof of Theorem \ref{0327:th}]
Skewcommutativity of the $\lambda$-bracket \eqref{0327:eq9} is clear, 
since, for odd $N$, the operators $H^{(N,0)}(D)$ is skewadjoint.
We thus only have prove the Jacobi identity:
\begin{equation}\label{0327:pm1}
\big\{u_\lambda\{u_\mu u\}\big\}
- \big\{u_\mu\{u_\lambda u\}\big\}
=\big\{\{u_\lambda u\}_{\lambda+\mu}u\big\}\,.
\end{equation}
By Lemma \ref{0327:lem2}, the first term in the LHS of \eqref{0327:pm1} is, 
denoting $M=m+3,\,N=n+3$,
\begin{equation}\label{0327:pm2}
\begin{array}{l}
\displaystyle{
\sum_{M,N} \alpha_M\alpha_N \big\{u_\lambda\{u_\mu u\}_N\big\}_M
= \sum_{M,N} \alpha_M\alpha_N (\lambda+\mu+D)^2
\big\{u_\lambda B^{(n)}(\mu+D)\mu\big\}_M 
}\\
\displaystyle{
= \sum_{M,N} \alpha_M\alpha_N (\lambda+\mu+D)^2
\bigg(
\frac1u
\Big((\lambda+D)  B^{(m)}(\lambda+D)\lambda\Big)
\Big((\mu+D)  B^{(n)}(\mu+D)\mu\Big) 
}\\
\displaystyle{
\,\,\,\,\,\,\,\,\,\,\,\,\,\,\,\,\,\,\,\,\,\,\,\,\,\,\,\,\,\,\,\,\,\,\,\,\,\,\,\,\,\,\,\,\,\,\,\,\,\,\,
-\lambda^2\mu^2
B^{(n)}(\lambda+\mu+D)  B^{(m)}(\lambda+D)\frac1u
\bigg)\,.
}
\end{array}
\end{equation}
The first term in the RHS of \eqref{0327:pm2} is invariant under the exchange of $\lambda$ and $\mu$,
hence it gives no contribution to the Jacobi identity.
Hence, the LHS of \eqref{0327:pm1} is
\begin{equation}\label{0327:pm3}
-\lambda^2\mu^2 \sum_{M,N} \alpha_M\alpha_N (\lambda+\mu+D)^2 
B^{(n)}(\lambda+\mu+D) 
\bigg(
B^{(m)}(\lambda+D)\frac1u - B^{(m)}(\mu+D)\frac1u
\bigg)\,.
\end{equation}
By Lemma \ref{0327:lem1}, the RHS of \eqref{0327:pm1} is
$$
\begin{array}{l}
\displaystyle{
\sum_{M,N} \alpha_M\alpha_N
\big\{{\{u_\lambda u\}_M}_{\lambda+\mu}u\big\}_N
=
\sum_{M,N} \alpha_M\alpha_N \lambda \mu^2
\big\{{B^{(m)}(\lambda+D)1}_{\lambda+\mu}u\big\}_N 
}\\
\displaystyle{
=
- \lambda^2 \mu^2 \sum_{M,N} \alpha_M\alpha_N 
(\lambda+\mu+D)^2  B^{(n)}(\lambda+\mu+D) 
\Big(
B^{(m)}(\lambda+D)\frac1u
-B^{(m)}(\mu+D)\frac1u
\Big)\,,
}
\end{array}
$$
which is equal to \eqref{0327:pm3}.
\end{proof}



\begin{thebibliography}{999999}
\bibitem[A]{A} A.M. Astashov, {\it Normal forms of Hamiltonian operators in 
field theory}, 
Soviet Math. Doklady \textbf{270} (1983), 1033-1037 (in Russian).


\bibitem[AV]{AV} A.M. Astashov and A.M. Vinogradov, 
{\it On the structure of Hamiltonian operators in field theory}, 
J. Geom. Phys. \textbf{2} (1986), 263-287 (in Russian).

\bibitem[BD]{BD} A. Beilinson and V.G. Drinfeld, {\it Chiral algebras}, AMS
Colloquium Publications, vol 51, Amer. Math. Soc., Providence, RI, 2004.

\bibitem[BDK]{BDK} A. Barakat, A. De Sole and V.G. Kac, {\it Poisson vertex
algebras in the theory of Hamiltonian equations}, Japan. J. Math. \textbf{4} 
(2009), 141--252.  

\bibitem[C]{C} D. Cooke, {\it Classification results and Darboux' theorem 
for low order Hamiltonian operators}, J. Math.Phys. \textbf{32}(1991), 109-119.

\bibitem[C1]{C1} D. Cooke, {\it Compatibility conditions for Hamiltonian 
pairs}, J. Math.Phys. \textbf{32}(1991), 3071-3076.

\bibitem[D]{D} I. Dorfman, {\it Dirac structures and integrability of 
non-linear evolution equations}, John Wiley and sons, New York, 1993. 

\bibitem[DK]{DK} A. De Sole and V.G. Kac, {\it Finite vs affine $W$-algebras},
Japan. J. Math. \textbf{1} (2006), 137--261. 

\bibitem[GD]{GD} I.M. Gelfand and I. Dorfman, {\it Hamiltonian operators and
algebraic structures related to them}, Funct. Anal. Appl. \textbf{13}
(1979),248-262.
 
\bibitem[FT]{FT} L.D. Faddeev and L.A. Takhtajan, {\it Hamiltonian approach
in soliton theory}, Nauka, 1986.

\bibitem[K]{K} V. G. Kac, {\it Vertex algebras for beginners}, Univ. Lecture 
Ser. vol 10, Amer. Math. Soc., Providence, RI, 1996. Second edition 1998. 

\bibitem[Ma]{Ma} F. Magri, {\it A simple model of the integrable Hamiltonian
equation}, J. Math. Phys. \textbf{19} (1978), 1156-1162.

\bibitem[MSS]{MSS} A.V. Mikhailov, A.V. Shabat and V.V. Sokolov {\it
The symmetry approach to classification of integrable equations},
in: V.E. Zakharov, ed., What is integrabilty?, Springer series in
non-linear dynamics, 1991, 115-184.
 
\bibitem[M]{M} O.I. Mokhov, {\it Hamiltonian differential operators and 
contact geometry}, Funct. Anal. Appl. \textbf{21:3} (1987), 53-60 (in Russian).

\bibitem[O]{O} P.J. Olver, {\it Darboux' theorem for Hamiltonian operators}, 
J. Diff. Eq. \textbf{71} (1988), 10-33.


\bibitem[V]{V} A.M. Vinogradov, \it {On Hamiltonian structures in 
field theory}, Sov. Math. Doklady \textbf{241} (1978), 18-21 (in Russian).

\end{thebibliography}
\end{document}